\documentclass[english,final,table]{IEEEtran}
\usepackage[T1]{fontenc}
\usepackage[latin9]{inputenc}
\usepackage{xcolor}
\usepackage{float}
\usepackage{mathrsfs}
\usepackage{bm}
\usepackage{amsmath}
\usepackage{amsthm}
\usepackage{amssymb}
\usepackage{graphicx}
\PassOptionsToPackage{normalem}{ulem}
\usepackage{ulem}

\makeatletter

\floatstyle{ruled}
\newfloat{algorithm}{tbp}{loa}
\providecommand{\algorithmname}{Algorithm}
\floatname{algorithm}{\protect\algorithmname}
\providecolor{lyxadded}{rgb}{0,0,1}
\providecolor{lyxdeleted}{rgb}{1,0,0}

\DeclareRobustCommand{\lyxsout}[1]{\ifx\\#1\else\sout{#1}\fi}

\usepackage{amsmath,amssymb}
\usepackage[level=0]{wgroup_message}  

\usepackage{color}  
\usepackage{graphicx,psfrag,cite,subfigure}

\usepackage{flushend}

\author{
Wangqian~Chen
and Junting~Chen

\thanks{The work was supported in part by the National Science Foundation of China (NSFC)
under Grant No. 62293482, by the Basic Research Project No. HZQB-KCZYZ-2021067 of
Hetao Shenzhen-HK S\&T Cooperation Zone, by NSFC Grant No. 62171398,
by the Shenzhen Science and Technology Program under
Grant No. JCYJ20210324134612033 and No. KQTD20200909114730003,
by Guangdong Research Projects No. 2019QN01X895, No. 2017ZT07X152,
and No. 2019CX01X104, by the Shenzhen Outstanding Talents Training Fund 202002,
by the Guangdong Provincial Key Laboratory of Future Networks of
Intelligence (Grant No. 2022B1212010001), by the National Key R\&D Program of China
with grant No. 2018YFB1800800, and by the Key Area R\&D Program of Guangdong Province
with grant No. 2018B030338001.
}

\thanks{W.~Chen and J.~Chen are with the School of Science and Engineering,
and the Future Network of Intelligence Institute (FNii), The Chinese University of Hong Kong, Shenzhen,
Guangdong 518172, China (email: wangqianchen@link.cuhk.edu.cn; juntingc@cuhk.edu.cn).}

}


\usepackage[acronym]{glossaries}
\newcommand{\newac}{\newacronym}
\newcommand{\ac}{\gls}

\makeglossaries
\newac{speb}{SPEB}{square position error bound}
\newac[plural=EFIMs,firstplural=Fisher information matrices (EFIMs)]{efim}{EFIM}{Fisher information matrix}
\newac{ne}{NE}{Nash equilibrium}
\newac{mse}{MSE}{mean squared error}
\newac{toa}{TOA}{time-of-arrival}
\newac{snr}{SNR}{signal-to-noise ratio}
\newac{lan}{LAN}{local area network}
\newac{psd}{PSD}{positive semidefinite}
\newac{pd}{PD}{positive definite}
\newac{wrt}{w.r.t.}{with respect to}
\newac{lhs}{L.H.S.}{left hand side}
\newac{wp1}{w.p.1}{with probability 1}
\newac{kkt}{KKT}{Karush-Kuhn-Tucker}
\newac{wlog}{w.l.o.g.}{without loss of generality}
\newac{mle}{MLE}{maximum likelihood estimation}
\newac{gps}{GPS}{global positioning system}
\newac{rssi}{RSSI}{received signal strength indicator}
\newac{mimo}{MIMO}{multiple-input multiple-output}
\newac{csi}{CSI}{channel state information}
\newac{fdd}{FDD}{frequency division duplexing}
\newac{ms}{MS}{mobile station}
\newac{bs}{BS}{base station}
\newac{d2d}{D2D}{device-to-device}
\newac{slnr}{SLNR}{signal-to-interference-leakage-and-noise-ratio}
\newac{ula}{ULA}{uniform linear antenna array}
\newac{pas}{PAS}{power angular spectrum}
\newac{mmse}{MMSE}{minimum mean square error}
\newac{zf}{ZF}{zero-forcing}
\newac{rzf}{RZF}{regularized zero-forcing}
\newac{as}{AS}{angular spread}
\newac{aod}{AOD}{angle of departure}
\newac{iid}{i.i.d.}{independent and identically distributed} 
\newac{sinr}{SINR}{signal-to-interference-and-noise ratio}
\newac{tdd}{TDD}{time-division duplex}
\newac{rvq}{RVQ}{random vector quantization}
\newac{rhs}{R.H.S.}{right hand side}
\newac{mrc}{MRC}{maximum ratio combining}
\newac{cdf}{CDF}{cumulative distribution function}
\newac{a.s.}{a.s.}{almost surely}
\newac{los}{LOS}{line-of-sight}
\newac{jsdm}{JSDM}{joint spatial division and multiplexing}
\newac{map}{MAP}{maximum a posteriori}
\newac{klt}{KLT}{Karhunen-Lo\`eve Transform}
\newac{lbe}{LBE}{link bargaining equilibrium}
\newac{se}{SE}{Stackelberg equilibrium}
\newac{uav}{UAV}{unmanned aerial vehicle}
\newac{nlos}{NLOS}{non-line-of-sight}
\newac{pdf}{PDF}{probability density function}
\newac{em}{EM}{expectation-maximization}
\newac{knn}{KNN}{$k$-nearest neighbor}
\newac{svd}{SVD}{singular value decomposition}
\newac{nmf}{NMF}{non-negative matrix factorization}
\newac{umf}{UMF}{unimodality-constrained matrix factorization}
\newac{rmse}{RMSE}{rooted mean squared error}
\newac{olos}{OLOS}{obstructed line-of-sight}
\newac{mmw}{mmW}{millimeter wave}
\newac{ber}{BER}{bit error rate}
\newac{rss}{RSS}{received signal strength}
\newac{lp}{LP}{linear program}
\newac{ufw}{U-FW}{unimodal Frank-Wolfe}
\newac{utf}{UTF}{unimodality-constrained tensor factorization}
\newac{fw}{FW}{Frank-Wolfe}
\newac{iot}{IoT}{Internet-of-Things}
\newac{mae}{MAE}{mean absolute error}
\newac{crb}{CRB}{Cram\'er-Rao bound}
\newac{aoa}{AoA}{angle of arrival}
\newac{wcl}{WCL}{weighted centroid localization}

\usepackage{enumitem}

\usepackage{algorithmic}

\usepackage{multirow}
\usepackage{makecell}

\makeatother

\usepackage{babel}
\begin{document}
\title{Diffraction and Scattering Aware Radio Map and Environment Reconstruction
using Geometry Model-Assisted Deep Learning}
\maketitle
\begin{abstract}

Machine learning (ML) facilitates rapid channel modeling for 5G and beyond wireless communication systems. Many existing ML techniques utilize a city map to construct the radio map; however, an updated city map may not always be available. 
This paper proposes to employ the \ac{rss} data to jointly construct the radio map and the virtual environment by exploiting the geometry structure of the environment.
In contrast to many existing ML approaches that lack of an environment model, we develop a virtual obstacle model and characterize the geometry relation between the propagation paths and the virtual obstacles.
A multi-screen knife-edge model is adopted to extract the key diffraction features, and these features are fed into a neural network (NN) for diffraction representation. 
To describe the scattering, as oppose to most existing methods that directly input an entire city map, our model focuses on the geometry structure from the local area surrounding the TX-RX pair and the spatial invariance of such local geometry structure is exploited.
Numerical experiments demonstrate that, in addition to reconstructing a 3D virtual environment, the proposed model outperforms the state-of-the-art
methods in radio map construction with $10\%-$$18\%$ accuracy improvements. It can also reduce $20\%$ data and $50\%$ training epochs when transferred to a new environment.
\end{abstract}

\begin{IEEEkeywords}
Radio map, Environment sensing, Diffraction features, Scattering-aware,
Deep learning, Neural network
\end{IEEEkeywords}

\section{Introduction}

\IEEEPARstart{R}{adio} maps find extensive applications in wireless
communications, such as network planning, localization and cognitive
radio systems \cite{MuLiuGuo:J21,WuYanXia:J17,ZenXu:J21,HanXue:J20}.
For example, radio maps have been constructed and exploited to assist
for low altitude \ac{uav} communications \cite{ZhaShuRui:J21,MoHuaXu:J20,9354009,9119191},
in which radio maps can help \ac{uav} seek \ac{los} opportunities
to serve for the ground users that are probably blocked by obstacles
in a dense urban environment.

Classical radio map construction can be roughly categorized into ray
tracing (RT) approaches, interpolation approaches, and deep learning
based approaches. RT approaches \cite{ZhuMaoSon:J22,LimChoSim:J20,SugaSas:J21,SugaSas:J23}
construct the propagation channel by launching propagation rays via
geometry analysis based on 3D digital environment models. Fine-grained
propagation models including reflection models, diffraction models,
and scattering models as well as the electromagnetic coefficients
of the materials in the environment are required for evaluating the
channel coefficients. However, obtaining an accurate digital environment
map with sufficient details for an accurate RT is very challenging.
It is also computationally intense to simulate the propagation rays
due to the complexity of an actual environment. Moreover, based on
the classical RT models, it is almost prohibitive to solve the inverse
problem that reconstructs the geometry of the environment based on
the radio measurements.

Interpolation-based radio map construction techniques exploit real
measurements taken at various locations of transmitters (TXs) and
receivers (RXs) without explicitly exploiting the geometry structure
of the environment. Some representative interpolation methods for
radio map construction include $k$-nearest neighbor (KNN) interpolation
\cite{ZhaWan:J22}, inverse distance weighted (IDW) interpolation
\cite{PhiTonSic:C12}, matrix completion \cite{SunChen:22J}, dictionary-based
compressive sensing \cite{AusAndNev:J18}, and Kriging \cite{HuZha:J20},
etc. These methods are based on the spatial correlation of measurements,
but they cannot differentiate the corresponding propagation conditions,
such as \ac{los} and \ac{nlos}. To address this issue, a segmented
propagation model was proposed in \cite{CheYatGes:C17}, and the measurement
data is clustered into different categories corresponding to the different
propagation conditions using subspace clustering, where different
propagation models are applied locally for constructing radio map.
In general, many interpolation-type methods assume that the TX locations
are fixed. When the TX location also varies, conventional interpolation-type
methods may suffer from the curse of dimensionality issue, where a
prohibitive amount of data is needed to learn the spatial structure
of the radio map.

Deep learning approaches for radio map construction were attempted
in \cite{KenSaiChe:J19,PopJefAta:J19,WuDanAi:20J}, where the features
that are fed into the neural network (NN) include building heights,
propagation distances, and the environment maps. In \cite{LevYapKut:J21,JuOmeDhe:22C},
a convolutional neural network (CNN) based structure was proposed
to learn the channel path loss based on 3D city maps. Some studies
also proposed to fuse the information from satellite images to learn
the characteristics of the propagation environment \cite{AteHas:19J,ThrZibDar:20J,AhmOmaAte:20J}.
In addition, with an adequate number of radio measurements, deep completion
autoencoders \cite{TegRom:J21} and generative adversarial network
\cite{HuHua:J23} were also adopted for radio map construction.

To summarize, most machine learning approaches directly interpolate
the radio map from a large number of measurements or they fuse the
environment information to assist for the interpolation, without explicitly
exploiting the geometry structure of the environment. Although some
existing CNN based approaches employ city maps as input, the maps
are usually treated as images rather than 3D structure. As a result,
it is challenging to build a radio map with {\em full} spatial degrees
of freedom where {\em both} the TX position and the RX position can
be varying over the whole 3D space; instead, existing approaches usually
require at least one coordinate of the TX and RX locations to be fixed.
Moreover, little is known on how to reconstruct the propagation environment
from pure radio measurements when a city map is not available or inaccurate.

This paper attempts to construct a radio map with full spatial degrees
of freedom for the large-scale channel quality between variable TX
and RX locations based on received signal strength (RSS) measurements.
In contrast to many existing machine learning approaches that lack
of an environment model, we aim at constructing a propagation model
that explicitly characterizes the propagation attenuation as a function
of the {\em geometry structure} of the propagation environment. In
addition, we propose to {\em jointly} reconstruct the geometry of
the propagation environment and the radio map. Towards this end, we
develop a virtual obstacle model and we characterize the propagation
attenuation by analyzing the geometry relation between the propagation
rays and the virtual obstacles. While our earlier attempts \cite{LiuChen:23J,ZenChe:C22}
only considered signal attenuation due to blockage along the direct
path, some significant factors including diffraction and reflection
were ignored, resulting in an oversimplified propagation model. In
a related approach, simultaneous localization and mapping (SLAM) usually
assumes at most one reflection and no diffraction as a compromise
with the complexity. In this work, we attempt to also model the effect
of diffraction and scattering for a more accurate radio map construction
as well as a virtual environment reconstruction.

Specifically, we adopt the virtual obstacle concept developed in \cite{LiuChen:23J,ZenChe:C22}
, but the virtual obstacle constructed in this paper not only models
the signal blockage of the direct path, but also captures the signal
attenuation due to diffraction and scattering. A multi-screen knife-edge
model is adopted to extract the key diffraction parameters of the
propagation path, and these parameters are fed into an NN for diffraction
representation. To describe the scattering, as oppose to most existing
methods that directly map an entire city map or satellite image to
a radio map, our model only focuses on the geometry structure from
the {\em local area} surrounding the TX-RX position pair, resulting
in efficient data augmentation and model parameter reduction.

The novelty and contribution are summarized as follows:
\begin{itemize}
\item We propose a deep learning model to jointly construct a 6D radio map
and a virtual geometry of the propagation environment from \ac{rss}
measurements. The model consists of an LOS branch, a diffraction branch,
and a scattering branch that explicitly or implicitly characterize
the geometry relation between the propagation attenuation and the
virtual environment.
\item In the diffraction branch, we adopt a transformer structure to approximate
the Vogler's method for an efficient computation of the diffraction
coefficient. In the scattering branch, we develop a rotation invariant
and scale invariant mapping for data augmentation to learn the geometry
structure of the local scatters, enhancing the convergence of the
CNN model parameters.
\item Simulations demonstrate that the proposed model outperforms the state-of-the-art
methods in radio map construction with $10\%-$$18\%$ accuracy improvements
and it can be generalized well to the whole 3D space without any additional
data to fine-tune the parameters.
\item Transferability of the proposed model is demonstrated with nearly
$20\%$ data reduction and $50\%$ training epoch reduction compared
with the non-transferred cases.
\item We present an application of radio-map-based UAV relay placement in
which the proposed model accelerates the UAV relay positioning with
reducing 99\% search distance with the classical radio-map-based methods.
\end{itemize}

The rest of the paper is organized as follows. The system model are
reviewed in Section II. Section III discusses the proposed learning
framework. Design examples are presented in Section IV and conclusions
are drawn in Section V.

\section{System Model}

Denote a wireless channel $\tilde{\mathbf{p}}=(\mathbf{p}_{\mathrm{t}},\mathbf{p}_{\mathrm{r}})\in\mathbb{R}^{6}$
using the positions $\mathbf{p}_{\mathrm{t}},\mathbf{p}_{\mathrm{r}}\in\mathbb{R}^{3}$
of the TX and RX, respectively. Consider a blockage-aware channel
model that describes the channel attenuation in logarithmic scale
for $\tilde{\mathbf{p}}$ as
\begin{equation}
g_{\mathrm{[dB]}}(\tilde{\mathbf{p}},\mathbf{H})=g_{\mathrm{0}}(\tilde{\mathbf{p}},\mathbf{H})+g_{\mathrm{d}}(\tilde{\mathbf{p}},\mathbf{H})+g_{\mathrm{s}}(\tilde{\mathbf{p}},\mathbf{H})+\xi\text{(\ensuremath{\tilde{\mathbf{p}}})}\label{eq:radio-map-model}
\end{equation}
where $\mathbf{H}$ is a parameter to describe the propagation environment.
The first term $g_{\mathrm{0}}(\tilde{\mathbf{p}},\mathbf{H})$ captures
the path loss if the link $\tilde{\mathbf{p}}$ is in the \ac{los}
condition, the second term $g_{\mathrm{d}}(\tilde{\mathbf{p}},\mathbf{H})$
describes the additional loss due to diffraction when the link is
in \ac{nlos} condition, the third term $g_{\mathrm{s}}(\tilde{\mathbf{p}},\mathbf{H})$
describes the fluctuation of the attenuation due to the other scattering
effects, and finally, the term $\xi(\tilde{\mathbf{p}})$ is a random
component that captures the uncertainty due to the small-scale fading.

The radio map studied in this paper captures the large-scale channel
information for each link $\tilde{\mathbf{p}}$ in the area of interest,
i.e.,
\begin{equation}
G(\tilde{\mathbf{p}})=g_{\mathrm{0}}(\tilde{\mathbf{p}},\mathbf{H})+g_{\mathrm{d}}(\tilde{\mathbf{p}},\mathbf{H})+g_{\mathrm{s}}(\tilde{\mathbf{p}},\mathbf{H}).\label{eq:radio-map-model-1}
\end{equation}
Note that the environment parameter $\mathbf{H}$ is a hidden variable
that can be learned from the data.

We measure only the location-labeled \ac{rss} $y(\tilde{\mathbf{p}})=G(\tilde{\mathbf{p}})+n$
at various locations $\tilde{\mathbf{p}}$ for the construction of
the radio map, where the small-scale fading $\xi(\tilde{\mathbf{p}})$
in (\ref{eq:radio-map-model}) is assumed to be partially averaged
and absorbed as the measurement noise $n$.

\subsection{Blockage Model}

We adopt a virtual obstacle model to describe the propagation environment.
Partition the ground area of interest into $M$ grid cells. The virtual
obstacle on the $m$th grid cell is modeled as a cube that occupies
the $m$th grid with height $h_{m}$ as shown in Fig.~\ref{fig:Geo-Dif}.
Denote $\mathbf{h}$ as a vector for the collection of all $h_{m}$,
$m=1,2,\dots,M$, and denote $\mathbf{H}$ as the matrix form of $h_{m}$
such that $\mathbf{h}=\mbox{vec}(\mathbf{H})$, where the $(i,j)$th
entry of $\mathbf{H}$ represents the virtual obstacle at the $(i,j)$th
grid cell. Hence, the variable $\mathbf{H}$ completely characterizes
the geometry of the virtual environment. Note that these virtual obstacles
do not have to exactly match with the city map, but they are to be
fitted such that the channel quality based on the radio map model
(\ref{eq:radio-map-model-1}) matches with the radio measurements.

For each position pair $\tilde{\mathbf{p}}$, denote $\mathcal{B}(\tilde{\mathbf{p}})$
as the set of grid cells that are covered by the line segment joining
$\mathbf{p}_{\mathrm{t}}$ and $\mathbf{p}_{\mathrm{r}}$. For each
grid cell $m\in\mathcal{B}(\tilde{\mathbf{p}})$, denote $z_{m}(\tilde{\mathbf{p}})$
as the height of the line segment of $\tilde{\mathbf{p}}$ that passes
over the $m$th grid cell. Denote $\mathcal{\tilde{D}}_{0}$ as the
set of channels $\tilde{\mathbf{p}}$ which are not blocked by any
virtual obstacle, and $\mathcal{\tilde{D}}_{0}$ is termed as the
(virtual) {\em LOS region} in this paper. Mathematically, for an
LOS case $\tilde{\mathbf{p}}\in\mathcal{\tilde{D}}_{0}$, we have
$h_{m}<z_{m}(\tilde{\mathbf{p}})$ for all grid cells $m\in\mathcal{B}(\tilde{\mathbf{p}})$;
for an NLOS case $\tilde{\mathbf{p}}\notin\mathcal{\tilde{D}}_{0}$,
we have $h_{m}\geq z_{m}(\tilde{\mathbf{p}})$ for at least one grid
cell $m\in\mathcal{B}(\tilde{\mathbf{p}})$, \emph{i.e.}, the direct
path from $\mathbf{p}_{\mathrm{t}}$ to $\mathbf{p}_{\mathrm{r}}$
is blocked by at least one of the virtual obstacles along the path.
Thus, a model for the LOS region $\tilde{\mathcal{D}}_{0}(\mathbf{H})$
given the parameter $\mathbf{H}$ for the heights of the virtual obstacles
can be formulated as
\begin{equation}
\begin{split}\mathbb{I}\{\tilde{\mathbf{p}}\in\mathcal{\tilde{D}}_{0}(\mathbf{H})\}=\prod_{m\in\mathcal{B}(\tilde{\mathbf{p}})}\mathbb{I}\{h_{m}<z_{m}(\tilde{\mathbf{p}})\}\end{split}
\label{eq:virtual-obstacle-model}
\end{equation}
where $\mathbb{I}\{A\}$ is an indicator function that takes value
$1$ if condition $A$ is true, and $0$, otherwise.

As a result, when a communication link is not blocked by any obstacle,
the channel attenuation is modeled as
\begin{equation}
\begin{split}g_{\mathrm{0}}(\tilde{\mathbf{p}},\mathbf{H})=(\beta_{0}+\gamma_{0}\log_{10}\|\mathbf{p}_{\mathrm{t}}-\mathbf{p}_{\mathrm{r}}\|_{2}\end{split}
)\mathbb{I}\{\tilde{\mathbf{p}}\in\tilde{\mathcal{D}}_{0}(\mathbf{H})\}\label{eq:LOS-component}
\end{equation}
where $\beta_{0}$ and $\gamma_{0}$ are the path loss parameters
to be learned for $\tilde{\mathbf{p}}\in\tilde{\mathcal{D}}_{0}$,
and $\|\cdot\|_{2}$ represents the Euclidean norm.

\subsection{Diffraction Model}

In propagation theory, when the direct path is blocked, the diffraction
mechanism allows the electromagnetic wave to propagate around obstacles
and contribute to the RX located in the shadow region. We adopt a
multi-screen knife-edge model reference, which assumes that the diffraction
effect mainly occurs on the edge of the obstacles, and the obstacles
are simplified as knife-edge shapes. We thus build a connection between
the diffraction and the virtual environment $\mathbf{H}$.

\begin{figure}[!t]
\centering \includegraphics[scale=0.3]{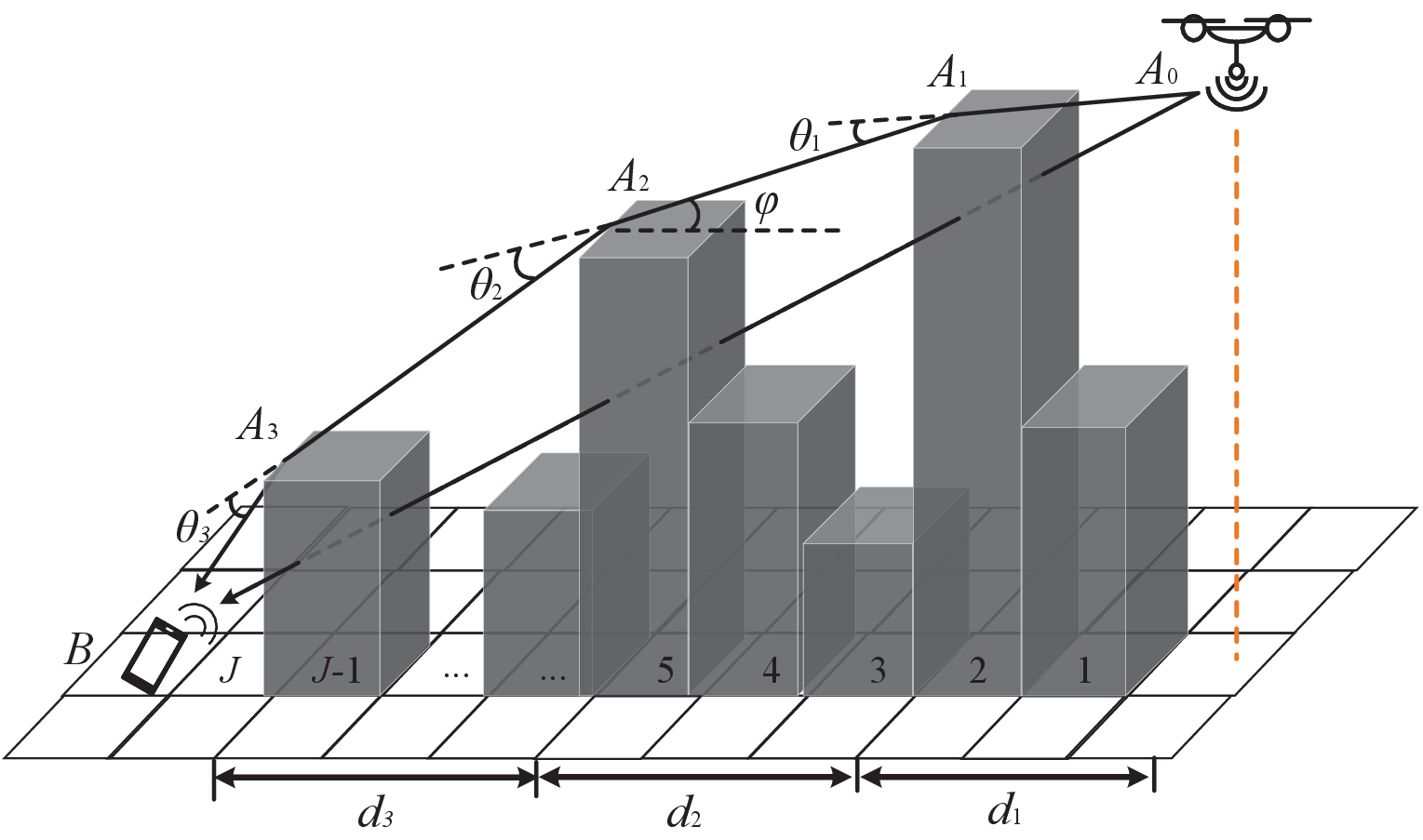}
\caption{Modeling the diffraction between point $A_{0}$ and point $B$ by
exploiting the geometry relation of distances $d_{i}$ and angles
$\theta_{i}$ from the virtual obstacles (gray cubes).}
\label{fig:Geo-Dif}
\end{figure}

Specifically, as illustrated in the example shown in Fig.~\ref{fig:Geo-Dif},
consider the set of grid cells $\mathcal{B}(\tilde{\mathbf{p}})$.
When the line segment $\overline{A_{0}B}$ joining $\mathbf{p}_{\mathrm{t}}$
and $\mathbf{p}_{\mathrm{r}}$ intersects with at least one virtual
obstacle, i.e., equation (\ref{eq:virtual-obstacle-model}) equals
to 0, then construct a concave piecewise linear curve $\overline{A_{0}A_{1}\cdots A_{3}B}$
with the shortest length, such that each vertex $A_{1},A_{2},\dots A_{N}$
touches the top of a virtual obstacle in $\mathcal{B}(\tilde{\mathbf{p}})$
and no segment $\overline{A_{i}A_{i+1}}$ intersects with any virtual
obstacles. The virtual obstacles where the vertexes of the curve touch
on are modeled as the multiple knife edges that mainly contribute
to the diffraction attenuation along the direct path, while all the
other obstacles below the height of the segment $\overline{A_{i}A_{i+1}}$
are neglected. The curve represents the main diffraction propagation
path around the obstacles from $\mathbf{p}_{\mathrm{t}}$ to $\mathbf{p}_{\mathrm{r}}$.

Define $d_{i}$ as the horizontal (ground) distance between vertices
$A_{i}$ and $A_{i-1}$, and $\theta_{i}$ as the diffraction angle
between adjacent line segments $\overline{A_{i}A_{i-1}}$ and $\overline{A_{i+1}A_{i}}$.
Vogler's method \cite{NguPhaMan:J21} suggests that the distances
$d_{i}$ and the angle $\theta_{i}$ in each segment are the key parameters
that affect the diffraction attenuation. See Appendix \ref{app:Vogler_method}
for a specific expression reported in the literature. The diffraction
effect can be modeled as a function $f_{\mathrm{d}}(\bm{d},\bm{\theta})$
that only depends on the distance variable $\bm{d}(\tilde{\mathbf{p}},\mathbf{H})=(d_{1},d_{2},\dots,d_{N})$
and the angle variable $\bm{\theta}(\tilde{\mathbf{p}},\mathbf{H})=(\theta_{1},\theta_{2},\dots,\theta_{N})$.
Considering that diffraction mainly occurs under the NLOS condition,
the diffraction component in the proposed radio map model (\ref{eq:radio-map-model-1})
is expressed as
\begin{equation}
\begin{split}g_{\mathrm{d}}(\tilde{\mathbf{p}},\mathbf{H})=f_{\mathrm{d}}(\bm{d},\bm{\theta})(1-\mathbb{I}\{\tilde{\mathbf{p}}\in\tilde{\mathcal{D}}_{0}(\mathbf{H})\}).\end{split}
\label{eq:diffeaction-model-NN}
\end{equation}

\subsection{Scattering Model}

We adopt an ellipse model that only focuses on the local scatters
in an elliptical area in the neighborhood of the ground positions
of $\mathbf{p}_{\mathrm{t}}$ and $\mathbf{p}_{\mathrm{r}}$. This
is because remote scatters are likely blocked by other obstacles and
the corresponding propagation distance is also longer, resulting in
a relatively small impact on the channel quality. In addition, to
compromise with the modeling complexity, we need the following properties
for our scattering model:
\begin{itemize}
\item {\em Rotation invariance}: Since only the relative positions of TX
and RX in the virtual environment matter with the propagation, the
same attenuation can be observed if one rotates the entire environment.
\item {\em Scale invariance}: Recall that the first two terms in the radio
map model (\ref{eq:radio-map-model-1}) depends on the propagation
distance. As the scattering component $g_{\mathrm{s}}(\tilde{\mathbf{p}},\mathbf{H})$
adds to the first two terms in log-scale, it aims at capturing the
relative attenuation. Thus, a distance-independent modeling may yield
a good approximation.
\end{itemize}

\begin{figure}[!t]
\centering \hspace{0.8cm}\includegraphics[scale=0.55]{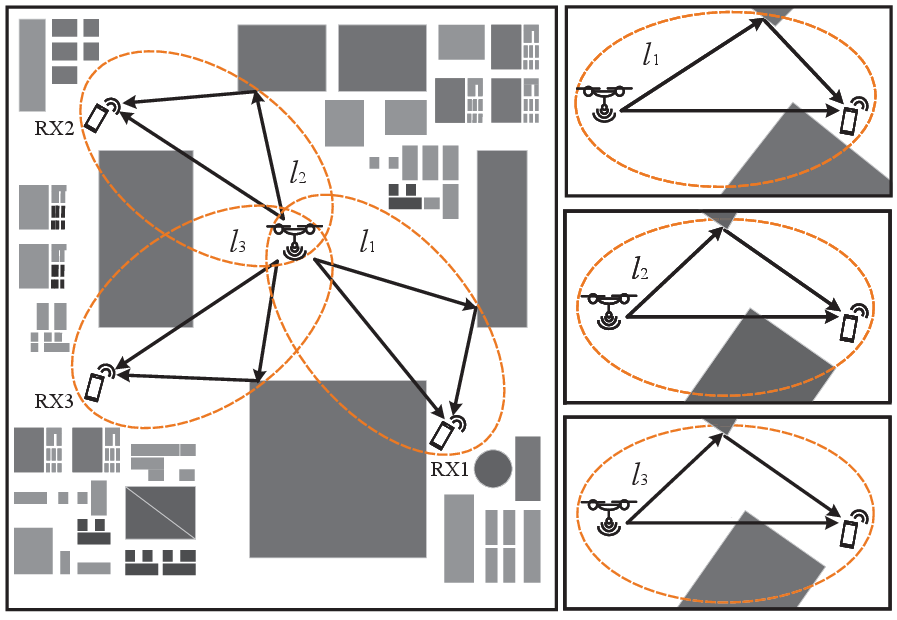}\caption{While the propagation environments from the TX to three RXs are different
(left figure), the local geometry structures are similar after proper
rotation and scaling (right figures). Thus, the same propagation mechanism
should be learned.}
\label{fig:Propagation-cases}
\end{figure}

This philosophy is demonstrated in Fig.~\ref{fig:Propagation-cases},
where there are three wireless channels from one TX to three RXs at
different locations. As shown in Fig.~\ref{fig:Propagation-cases},
the three channels share a similar local propagation environment,
where each channel consists of a slightly obstructed direct path with
diffraction and a strong reflected path. As such, one would expect
that the three channels share the same scattering component in (\ref{eq:radio-map-model-1}).

Specifically, the scattering component $g_{\text{s}}(\tilde{\mathbf{p}},\mathbf{H})$
is constructed as follows. Recall our notation for the TX position
$\mathbf{p}_{\mathrm{t}}=(\bar{\mathbf{p}}_{\textrm{t}},p_{\textrm{t,3}})\in\mathbb{R}^{3}$,
where $\bar{\mathbf{p}}_{\textrm{t}}=(p_{\textrm{t,1}},p_{\textrm{t,2}})\in\mathbb{R}^{2}$
is the ground projected location of the TX and $p_{\textrm{t,3}}$
is the altitude; similar notation $\mathbf{p}_{\mathrm{r}}=(\bar{\mathbf{p}}_{\textrm{r}},p_{\textrm{r,3}})$
is defined for the RX. Denote $d_{0}=\|\bar{\mathbf{p}}_{\textrm{t}}-\bar{\mathbf{p}}_{\textrm{r}}\|_{2}$
as the ground distance. Define an elliptical area using $\bar{\mathbf{p}}_{\textrm{t}}$
and $\bar{\mathbf{p}}_{\textrm{r}}$ as the foci with a given eccentricity
parameter $e$. As a result, the major and minor axes of the ellipse
are given by $d_{0}/e$ and $(d_{0}/e)\sqrt{1-e^{2}}$, respectively.
Define $\mathcal{B}_{\mbox{\scriptsize E}}(\tilde{\mathbf{p}},\mathbf{H})$
as the set of grid cells within the elliptical area. The scattering
component in (\ref{eq:radio-map-model-1}) is modeled as
\begin{equation}
\begin{split}g_{\mathrm{s}}(\tilde{\mathbf{p}},\mathbf{H})=f_{\text{s}}(\mathcal{B}_{\mbox{\scriptsize E}}(\tilde{\mathbf{p}},\mathbf{H}))(1-\mathbb{I}\{\tilde{\mathbf{p}}\in\tilde{\mathcal{D}}_{0}(\mathbf{H})\})\end{split}
\label{eq:scattering-model-NN}
\end{equation}
where $f_{\text{s}}(\mathcal{B}_{\mbox{\scriptsize E}}(\tilde{\mathbf{p}},\mathbf{H}))$
is a mapping that depends on a subset of entries of $\mathbf{H}$
based on $\mathcal{B}_{\mbox{\scriptsize E}}(\tilde{\mathbf{p}},\mathbf{H})$
and such a mapping $f_{\text{s}}$ satisfies the rotation invariance
and scale invariance properties.

\begin{figure*}[!t]
\centering \includegraphics[scale=0.6]{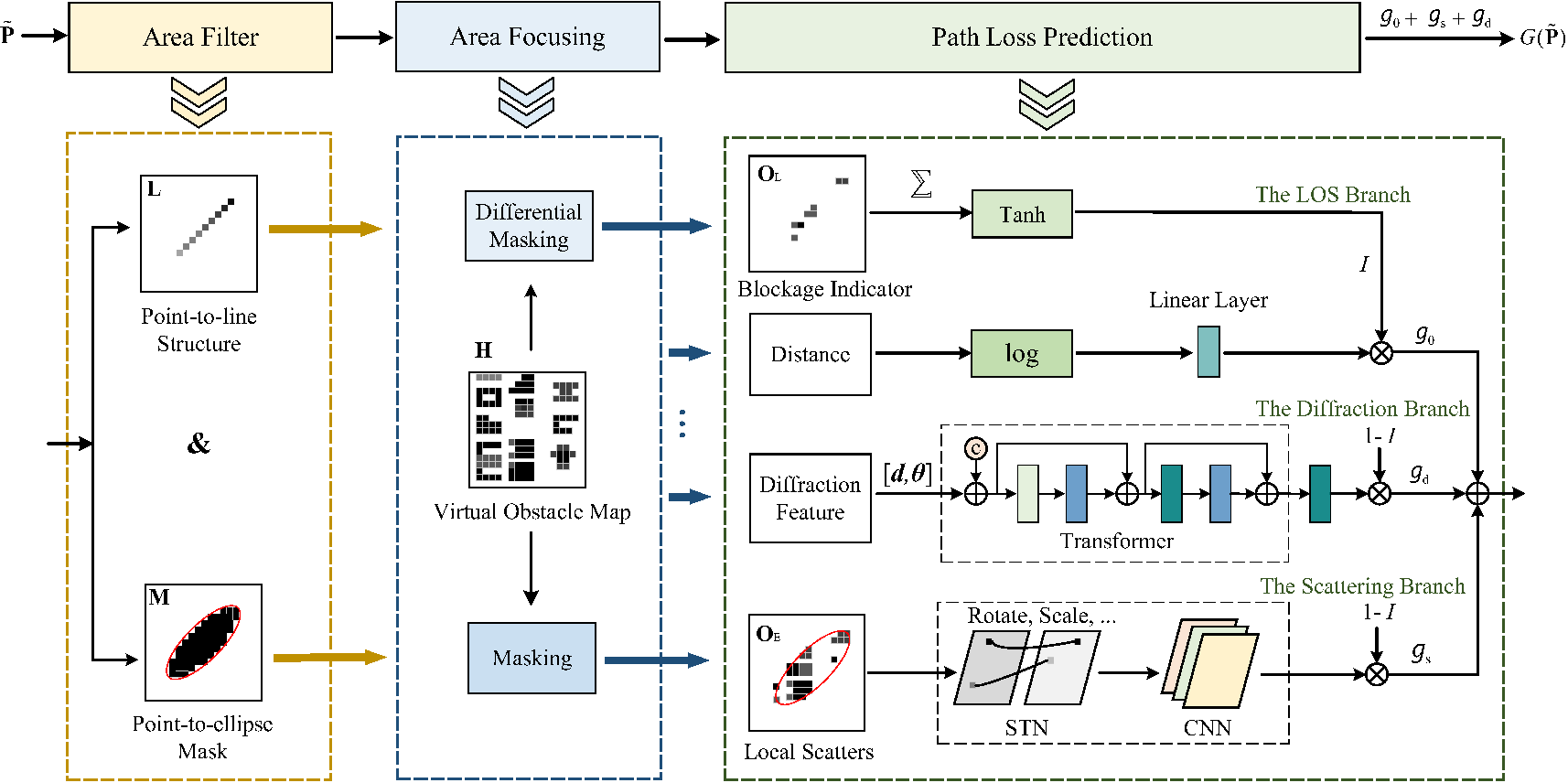}
\caption{The overall architecture to exploit the geometry structure of the
environment for joint 6D radio map and virtual environment map reconstruction.}
\label{fig:network-architecture}
\end{figure*}

\section{Network Architecture and Training}

In this section, a deep NN representation is constructed to model
the relationship between the radio map $G(\tilde{\mathbf{p}})$ and
the virtual obstacle environment $\mathbf{H}$. Specifically, we need
to address the following technical challenges:
\begin{itemize}
\item How to construct a deep learning architecture for (\ref{eq:radio-map-model-1})
such that the virtual obstacle map $\mathbf{H}$ can be efficiently
constructed from the \ac{rss} measurements? For example, it could
be challenging to find a NN representation for the blockage relation
(\ref{eq:virtual-obstacle-model}) such that the variable $\mathbf{H}$
can be efficiently updated by back-propagation for training.
\item How to construct a NN expression for the Vogler diffraction model
to exploit the geometry structure of diffraction? As is seen in Section
II-B and Appendix~\ref{app:Vogler_method}, the Vogler expression
has a special mathematical structure, and a conventional fully connected
NN (FCN) structure may not be a good choice for balancing the representation
capability and the learning efficiency.
\item How to design a CNN with the rotation invariance and scale invariance
properties for the scattering component? While CNN is a common structure
that can exploit both local and global features, it still needs to
extend its capability for the rotation and scale invariance.
\end{itemize}

In this section, we will explore specialized network structures, such
as multi-head attention and spatial transformation to address the
above issues. The overall architecture of the proposed network is
shown in Fig.~\ref{fig:network-architecture}.

\subsection{Overall Architecture}

As illustrated in Fig.~\ref{fig:network-architecture}, the proposed
network consists of three main modules, Area Filter, Area Focusing,
and Path Loss Prediction. For each module, there are two to three
branches focusing on the LOS component $g_{\mathrm{0}}$, the diffraction
component $g_{\mathrm{d}}$, and the scattering component $g_{\mathrm{s}}$
in (\ref{eq:radio-map-model-1}). The virtual obstacle environment
that has been modeled as a 2D virtual obstacle map $\mathbf{H}$ is
processed by the Area Focusing module.

\subsubsection{Area Filter}

Recall that the signal propagation for a wireless channel $\tilde{\mathbf{p}}$
is mostly determined by the local area surrounding the locations of
the TX $\mathbf{p}_{\text{t}}$ and the RX $\mathbf{p}_{\text{r}}$.
The Area Filter module thus aims at determining the relevant local
areas of $\tilde{\mathbf{p}}$ for the LOS component, the diffraction
component, and the scattering component.

Specifically, for a given $\tilde{\mathbf{p}}$, the Area Filters
are formed as $M_{1}\times M_{2}$ matrices to describe a 3D line
structure $\mathbf{L}$ and an ellipse mask $\mathbf{M}$, where the
3D line structure $\mathbf{L}$ aims at computing the level of obstruction
for the wireless channel $\tilde{\mathbf{p}}$ due to the virtual
obstacle map $\mathbf{H}$ and selecting the relevant virtual obstacles
$h_{m}$ to form the set $\mathcal{B}(\tilde{\mathbf{p}})$ in (\ref{eq:virtual-obstacle-model});
and the ellipse mask $\mathbf{M}$ aims at selecting the virtual obstacles
in the surrounding of $\tilde{\mathbf{p}}$ to construct $\mathcal{B}_{\mbox{\scriptsize E}}(\tilde{\mathbf{p}},\mathbf{H})$
and learn $f_{\text{s}}$ in (\ref{eq:scattering-model-NN}).

For the line structure, we project the altitude of the direct line
connecting $\mathbf{p}_{\text{t}}$ and $\mathbf{p}_{\text{r}}$ to
the ground. Recall that $\mathcal{B}(\tilde{\mathbf{p}})$ is the
set of grid cells where the line segment joining $\mathbf{p}_{\mathrm{t}}$
and $\mathbf{p}_{\mathrm{r}}$ passes through. Thus, for $m\in\mathcal{B}(\tilde{\mathbf{p}})$,
let $\bar{\mathbf{c}}_{m}\in\mathbb{R}^{2}$ be the corresponding
ground location of the $m$th grid. The $m$th element of the matrix
$\mathbf{L}$ is given by $L_{m}=(p_{\textrm{t,3}}-p_{\textrm{r,3}})\frac{\left\Vert \bar{\mathbf{c}}_{m}-\bar{\mathbf{p}}_{\textrm{r}}\right\Vert _{2}}{\left\Vert \bar{\mathbf{p}}_{\textrm{t}}-\bar{\mathbf{p}}_{\textrm{r}}\right\Vert _{2}}+p_{\textrm{r,3}}$,
if $m\in\mathcal{B}(\tilde{\mathbf{p}})$; and $L_{m}=0$, for $m\notin\mathcal{B}(\tilde{\mathbf{p}})$.

For the ellipse mask, the entry of the matrix $\mathbf{M}$ takes
value 1 if the entry is covered by the ellipse with an eccentricity
parameter $e$ and foci $\bar{\mathbf{p}}_{\text{t}}$ and $\bar{\mathbf{p}}_{\text{r}}$,
and it takes value 0 otherwise.

\subsubsection{Area Focusing}

The Area Focusing module is to select a subset of virtual obstacle
variables ${h_{m}}$ and form the local environment feature for the
wireless channel $\tilde{\mathbf{p}}$. As a result, only the subsect
of the relevant variables ${h_{m}}$ are updated through back-propagation
during the training process.

Based on the line structure $\mathbf{L}$, the line feature map of
Area Focusing can be formulated as
\begin{equation}
\begin{split}\mathscr{\mathbf{\mathscr{\mathbf{O}}}}_{\textrm{L}}=\textrm{ReLU}((\mathbf{H}-\mathbf{L})\odot\mathrm{\textrm{sign}}(\mathbf{L}))\end{split}
\label{eq:obstruction_feature-1}
\end{equation}
where $\textrm{ReLU}(\cdot)$ and $\textrm{sign}(\cdot)$ are element-wise
operators, with $\textrm{ReLU}(x)=\max(x,0)$ and $\textrm{sign}(x)=\mathbb{I}\{x>0\}$,
and $\odot$ is the Hadamard product. It can be verified that the
non-zero elements of $\mathscr{\mathbf{\mathscr{\mathbf{O}}}}_{\textrm{L}}$
indicate the locations where the propagation is blocked along the
direct path. In addition, the value of the non-zero entries of $\mathscr{\mathbf{\mathscr{\mathbf{O}}}}_{\textrm{L}}$
may represent the level of blockage.

Based on the ellipse mask, the ellipse feature map of Area Focusing
is calculated as
\begin{equation}
\begin{split}\mathscr{\mathbf{\mathscr{\mathbf{O}}}}_{\textrm{E}}=\mathbf{H}\odot\mathbf{M}\end{split}
\label{eq:obstruction_feature-1-1}
\end{equation}
where only the obstacle variable $h_{m}$ that locates inside the
ellipse indicated by the mask $\mathbf{M}$ are selected for $\mathbf{O}_{\text{E}}$.

\subsubsection{Path Loss Prediction}

The Path Loss Prediction module consists of three branches. As in
Fig.~\ref{fig:network-architecture}, the first two branches from
the top take the feature $\mathbf{O}_{\text{L}}$ to predict the path
loss from the LOS component and the diffraction component. The third
branch takes the feature $\mathbf{O}_{\text{E}}$ to predict the additional
attenuation from the scattering. These are illustrated as follows.

\subsection{The LOS Branch}

The LOS branch is to implement the indicator function in (\ref{eq:LOS-component}),
which is approximated by a $\mbox{tanh}$ function for a non-degenerated
gradient in the training phase. We thus have the following soft indicator
function as
\begin{equation}
\begin{split}I=1-\tanh(\textrm{sum}(\mathscr{\mathbf{\mathscr{\mathbf{O}}}}_{\textrm{L}}))\end{split}
\label{eq:obstruction_level}
\end{equation}
where $\textrm{sum}(\mathbf{A})=\sum_{ij}a_{ij}$ and $\tanh(x)=(e^{x}-e^{-x})/(e^{x}+e^{-x})$.

It follows that when the blockage indicator matrix $\mathbf{O}_{\text{L}}$
contains all zeros, implying that the propagation is not blocked by
any virtual obstacle, we have the soft indicator $I=1$. On the other
hand, when $\mathbf{O}_{\text{L}}$ contains non-zeros values and
$\textrm{sum}(\mathbf{O}_{\text{L}})\geq2$, we have $I\leq0.04$,
implying that the propagation is very likely blocked. Here, the 2-meter
margin is to match with the intuition from real measurements that
there is a soft transition from LOS to NLOS.

The path loss model in (\ref{eq:LOS-component}) is implemented by
a regression layer as seen from Fig.~\ref{fig:network-architecture}.

\subsection{The Diffraction Branch}

The diffraction branch is designed to mimic the computation structure
of the Vogler's method illustrated in Appendix~\ref{app:Vogler_method}.
The Vogler expression has a special recursive structure between the
computation of the $N$th diffraction $F_{N}$ and the $(N-1)$th
diffraction $F_{N-1}$. The diffraction distances $d_{i}$ and angles
$\theta_{i}$ play a key role in determining the diffraction attenuation.
However, such inherently recursive and sequential nature precludes
parallelization in practical implementation. The transformer\emph{
}network \cite{vaswani:C2017} processes all features using a multi-head
attention mechanism which models the dependencies of recursive structure
without the need of the order of time and make them highly parallelizable.
Therefore, we propose to design a transformer network to learn the
diffraction by exploiting the diffraction features of distances $d_{i}$
and angles $\theta_{i}$.

\subsubsection{Determine \textmd{$d_{i}$ and $\theta_{i}$}}

\begin{algorithm}
\begin{algorithmic}[1]

\STATE Input $\mathbf{p}_{\mathrm{t}}$, $\mathbf{p}_{\mathrm{r}}$,
$\mathbf{L}$ and $\mathbf{O}_{\text{L}}$

\STATE Generate indexes for grid cells in $\mathcal{B}(\tilde{\mathbf{p}})$
with $1,2,3,...,J$

\STATE Initialize $\mathcal{I}=\{\bar{\mathbf{p}}_{\textrm{t}}\}$,
$\mathcal{H}=\{p_{\textrm{t,3}}\}$, $\mathcal{S}=\mathcal{B}(\tilde{\mathbf{p}})$,
$l=0$, $h_{0}$$=p_{\textrm{t,3}}$ and $h_{J+1}$$=p_{\textrm{r,3}}$

\WHILE{($\mathcal{S}$ is not empty)}

\FOR{$j=l$+1 to $J+1$}

\STATE $\varphi_{j}=$ artan($\frac{h_{l}-h_{j}}{j}$)

\ENDFOR

\STATE $l=\{j\mid j=\textrm{argmin}(\varphi_{j}),\forall j>l\}$

\IF {$v=J+1$}

\STATE Add $\bar{\mathbf{r}}_{J+1}$ to $\mathcal{I}$ and add $h_{J+1}$
to $\mathcal{\mathcal{H}}$, go to step 15

\ELSE \STATE Add $\bar{\mathbf{r}}_{l}$ to $\mathcal{I}$ and add
$h_{l}$ to $\mathcal{\mathcal{H}}$

\ENDIF

\ENDWHILE

\FOR{$i=1$ to $\textrm{length}(\mathcal{I})-1$}

\STATE$d_{i}=\|\mathcal{I}_{i}-\mathcal{I}_{i+1}\|_{2}$ and

\begin{small}

$\begin{aligned}\theta_{i}= & \arctan\left(\frac{\mathcal{H}_{i}-\mathcal{H}_{i+1}}{\|\mathcal{I}_{i}-\mathcal{I}_{i+1}\|_{2}}\right)-\arctan\left(\frac{\mathcal{H}_{i-1}-\mathcal{H}_{i}}{\|\mathcal{I}_{i}-\mathcal{I}_{i-1}\|_{2}}\right)\end{aligned}
$

\end{small}

\ENDFOR

\STATE Output $d_{i}$ and $\theta_{i}$

\end{algorithmic}

\caption{Virtual obstacle selecting algorithm for determining diffraction distances
$d_{i}$ and angles $\theta_{i}$ .}

\label{Alg1:Feature-extract}
\end{algorithm}

Using the example depicted in Fig.~\ref{fig:Geo-Dif}, the curve
$\overline{A_{0}A_{1}\cdots A_{3}B}$ characterizes the diffraction
path and the distances $d_{i}$ and angles $\theta_{i}$ are related
to the locations of the vertexes. To locate the vertexes, order the
grid cells in $\mathcal{B}(\tilde{\mathbf{p}})$ as $1,2,...,J$ according
to the distance away from $\mathbf{p}_{\mathrm{t}}$. The coordinate
and height of the virtual obstacle in $j$th grid cell are given as
$\bar{\mathbf{r}}_{j}$ and $h_{j}$, respectively. To find the vertex
$A_{1}$, take $A_{0}$ as the base point and connect $A_{0}$ to
the vertexes of all the virtual obstacles in $\mathcal{B}(\tilde{\mathbf{p}})$.
Geometrically, the vertex $A_{1}$ is located in the $j$th grid cell
where the segment joining the vertex of the virtual obstacle and $A_{0}$
generates the smallest elevation angle at the virtual obstacle, which
is denoted as $\varphi$ in Fig.~\ref{fig:Geo-Dif}. More generally,
assume $A_{i}$ is obtained in the $K$th grid cell with $K<J$. To
search $A_{i+1}$, connect $A_{i}$ and the vertex of the virtual
obstacle in $j$th grid cell to generate the segments for all $j>K$,
and calculate the elevation angle $\varphi_{j}$ of each segment at
the virtual obstacle. The vertex $A_{i+1}$ is located in the $l$th
grid cell where $l=\{j\mid j=\textrm{argmin}(\varphi_{j}),\forall j>K\}$.
With these vertexes, the diffraction distances $d_{i}$ and angles
$\theta_{i}$ are determined. The details of the virtual obstacle
selecting algorithm for determining $d_{i}$ and $\theta_{i}$ are
summarized in Algorithm \ref{Alg1:Feature-extract}. The extracted
distances $d_{i}$ and angles $\theta_{i}$ are input into the transformer
network to learn the diffraction $f_{\mathrm{d}}(\bm{d},\bm{\theta})$.

\subsubsection{Transformer representation of diffraction}

\begin{figure}[!t]
\centering \includegraphics[scale=0.56]{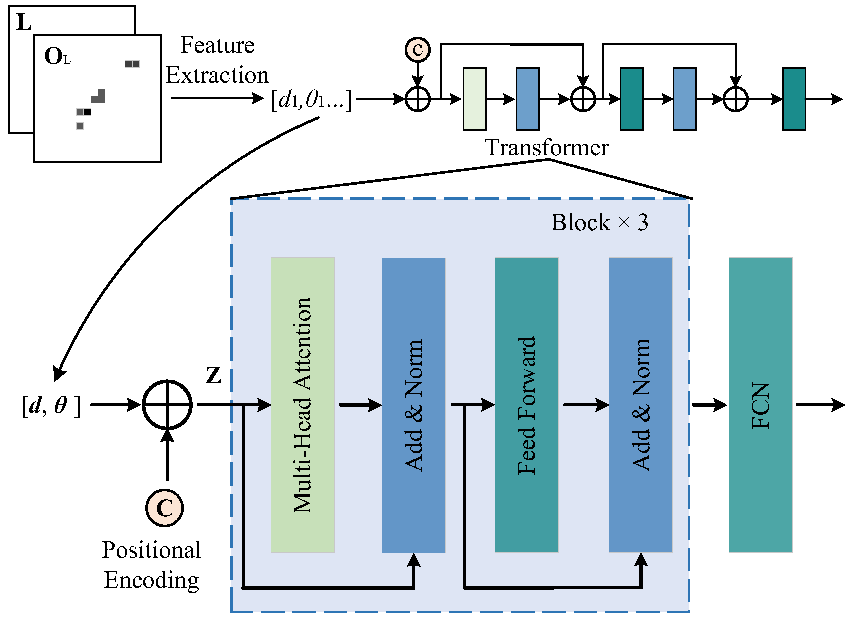}
\caption{The transformer to learn the diffraction mechanism by exploiting the
geometry structures of the diffraction distances $d_{i}$ and angles
$\theta_{i}$.}
\label{fig:FCN}
\end{figure}

The transformer network is shown in Fig.~\ref{fig:FCN}. The whole
network consists of a stack of three identical blocks followed with
a FCN to generate the diffraction component. Each block has a multi-head
attention layer, a feed forward layer and two normalization layers.
Residual connection is employed after the layer normalization. To
explore the sequential features of the distances $d_{i}$ and angles
$\theta_{i}$, positional encoding is implemented as an additional
input associated with the distances $d_{i}$ and angles $\theta_{i}$,
which are all input to an attention layer to learn their dependencies
between different positions to approximate the computation structure
of the Vogler expression.

\begin{figure*}[!t]
\centering \includegraphics[scale=0.5]{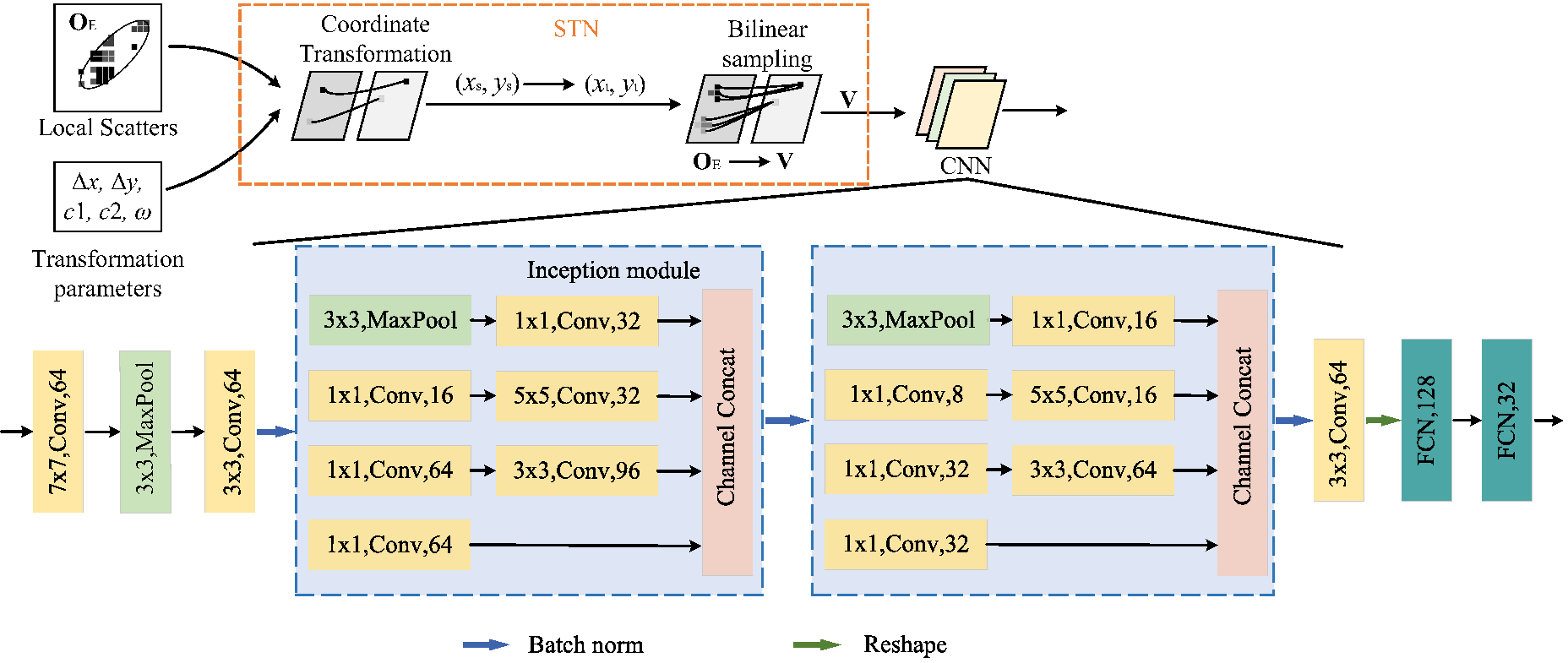} \caption{The CNN to learn the geometry structure of the local environment by
exploiting the rotation invariance and scale invariance properties.}
\label{fig:CNN}
\end{figure*}

\subsection{The Scattering Branch}

The scattering branch consists of a spatial transformation network
(STN) to capture the rotation and scaling invariance and a CNN to
learn the geometry structure of the local scatters. In STN, the local
scatters in the feature map of Area Focusing $\mathscr{\mathbf{\mathscr{\mathbf{O}}}}_{\textrm{E}}$
will be translated, rotated and scaled under a definitive rule. The
relative position information with geometry structure of the local
scatters is extracted. Followed with such operations, a CNN takes
these features of the local geometry structures to predict the scattering
component.

The STN is implemented by coordinate transformation and sampling \cite{JadSimZis:C15}.

\subsubsection{Coordinate Transformation}

Define the output feature map as $\mathbf{F}\in\mathbb{R}^{M_{1}\times M_{2}}$
that is translated, rotated and scaled from the input feature map
$\mathscr{\mathbf{\mathscr{\mathbf{O}}}}_{\textrm{E}}$. Let $(x_{\textrm{t}}^{(i)},y_{\textrm{t}}^{(i)})$
be the $i$th pixel coordinates in the output feature map $\mathbf{F}$
for $i=1,2,...M_{1}M_{2}$. The corresponding source coordinates $(x_{\textrm{s}}^{(i)},y_{\textrm{s}}^{(i)})$
of the input feature map $\mathscr{\mathbf{\mathscr{\mathbf{O}}}}_{\textrm{E}}$
under transformation is formulated as \begin{equation}
\resizebox{.95\hsize}{!}{$
\left[\begin{array}{c} x_{\textrm{s}}^{(i)}\\ y_{\textrm{s}}^{(i)} \end{array}\right]=
\left[\begin{array}{cc}
c_{1}\cos(\omega) & -c_{2}\sin(\omega)\\
c_{1}\sin(\omega) & \,\,\,\,c_{2}\cos(\omega) \end{array}\right]
\left[\begin{array}{c} x_{\textrm{t}}^{(i)}\\ y_{\textrm{t}}^{(i)} \end{array}\right]+
\left[\begin{array}{c} \Delta x\\
\Delta y \end{array}\right]\label{eq:transformation-all}
$}
\end{equation}where $\Delta x$ and $\Delta y$ are translation parameters of $x$
and $y$ coordinates, $c_{1}$ and $c_{2}$ are the scaling factors,
and $\omega$ is the rotation angle. Since the mapping above is from
integer value to real value, most pixels in the output feature map
$\mathbf{F}$ cannot find their exact counterparts in the input feature
map $\mathscr{\mathbf{\mathscr{\mathbf{O}}}}_{\textrm{E}}$. Therefore,
a bilinear sampling kernel is applied for calculating the values of
these pixels.

\subsubsection{Bilinear Sampling}

Using bilinear sampling, the pixel value $F(x_{\textrm{t}}^{(i)},y_{\textrm{t}}^{(i)})$
in the coordinates $(x_{\textrm{t}}^{(i)},y_{\textrm{t}}^{(i)})$
of the output feature map $\mathbf{F}$ will be calculated by weighting
the nearest four neighbor pixels around $(x_{\textrm{s}}^{(i)},y_{\textrm{s}}^{(i)})$
in the input feature map $\mathscr{\mathbf{\mathscr{\mathbf{O}}}}_{\textrm{E}}$,
which is formulated as
\begin{equation}
F(x_{\textrm{t}}^{(i)},y_{\textrm{t}}^{(i)})=\sum_{n}^{M_{1}}\sum_{m}^{M_{2}}O_{\textrm{E}}(n,m)w(n,m,x_{\textrm{s}}^{(i)},y_{\textrm{s}}^{(i)})\label{eq:sampling-weigth-1}
\end{equation}
where $w(n,m,x_{\textrm{s}}^{(i)},y_{\textrm{s}}^{(i)})$ is the weight
coefficient and $O_{\textrm{E}}(n,m)$ is the pixel value of the input
feature map $\mathscr{\mathbf{\mathscr{\mathbf{O}}}}_{\textrm{E}}$
in the coordinates $(n,m)$. The weights of the four nearest pixels
calculated by the bilinear sampling are given as $w(n,m,x_{\textrm{s}}^{(i)},y_{\textrm{s}}^{(i)})=(1-|x_{\textrm{s}}^{(i)}-n|)(1-|y_{\textrm{s}}^{(i)}-m|)$.
For the other pixels, the weights are zeros. Geometrically, the weights
of these pixels are zeros where the coordinate distances away from
$(x_{\textrm{s}}^{(i)},y_{\textrm{s}}^{(i)})$ are larger than $1$.
The weight coefficient for any pixel in $(x_{\textrm{s}}^{(i)},y_{\textrm{s}}^{(i)})$
can be consistently reformulated as \begin{equation}
\resizebox{.99\hsize}{!}{$
w(n,m,x_{\textrm{s}}^{(i)},y_{\textrm{s}}^{(i)})=\max(0,1-|x_{\textrm{s}}^{(i)}-n|)\max(0,1-|y_{\textrm{s}}^{(i)}-m|).\label{eq:sampling-weigth}
$}
\end{equation}Using (\ref{eq:sampling-weigth}), a weight coefficient vector $\mathbf{S}_{i}\in\mathbb{R}^{1\times M_{1}M_{2}}$
for the $i$th pixel of the output feature map $\mathbf{F}$ can be
defined as \[
\resizebox{.99\hsize}{!}{$
\mathbf{S}_{i}=\left[w(1,1,x_{\textrm{s}}^{(i)},y_{\textrm{s}}^{(i)}),w(1,2,x_{\textrm{s}}^{(i)},y_{\textrm{s}}^{(i)}),...w(n,m,x_{\textrm{s}}^{(i)},y_{\textrm{s}}^{(i)})\right]
$}
\]and therefore the output feature map $\mathbf{F}$ can be consistently
sampled as
\begin{equation}
\begin{split}\textrm{vec}(\mathbf{F})=\end{split}
\left[\begin{array}{c}
\mathbf{S}_{1}\\
\mathbf{S}_{2}\\
\vdots\\
\mathbf{S}_{M_{1}M_{2}}
\end{array}\right]\textrm{vec}(\mathscr{\mathbf{\mathscr{\mathbf{O}}}}_{\textrm{E}})\label{eq:sampling-1-1}
\end{equation}
where $\textrm{vec}(\cdot)$ is the column vectorization operation.
Taking $\mathscr{\mathbf{\mathscr{\mathbf{O}}}}_{\textrm{E}}$ with
(\ref{eq:obstruction_feature-1-1}), the rotation and scaling properties
can be applied to the virtual obstacle map $\mathbf{H}$, which helps
reconstruct the geometry structure of the virtual environment. Using
(\ref{eq:sampling-1-1}), it is also possible to downsample or oversample
a feature map, as one can define the output size to be different to
the input size, which can adjust the feature map in different scales
flexibly for CNN layer.

\subsubsection{Parameter Generation}

The knowledge of how to transform each feature map is solved by the
positions of the TX and RX. For position pair $\tilde{\mathbf{p}}$,
consider the line segment connecting the ground position of the TX
and RX for reference. The line segment is translated to the center
of the feature map and rotated to the horizontal direction while keeping
the relative locations of all scatters. The whole feature map will
be scaled based on the length of the line segment. As a result, the
parameters represent translation, scaling and rotation can be computed,
which are formulated as \[
\left[\begin{array}{c} \Delta x\\ \Delta y \end{array}\right]=
\left[\begin{alignedat}{1}\frac{p_{\textrm{t,1}}+p_{\textrm{r,1}}-M_{1}}{M_{1}}\\
\frac{p_{\textrm{t,2}}+p_{\textrm{r,2}}-M_{2}}{M_{2}} \end{alignedat} \right],\left[\begin{array}{c} c_{1}\\
c_{2} \end{array}\right]=\left[\begin{alignedat}{1}\frac{\|\bar{\mathbf{p}}_{\textrm{t}}-\bar{\mathbf{p}}_{\textrm{r}}\|_{2}}{M_{1}}\\ \frac{\|\bar{\mathbf{p}}_{\textrm{t}}-\bar{\mathbf{p}}_{\textrm{r}}\|_{2}}{M_{2}} \end{alignedat} \right]
\]and $\omega=-\arctan((p_{\textrm{t,2}}-p_{\textrm{r,2}})/(p_{\textrm{t,1}}-p_{\textrm{r,1}}))$
if $p_{\textrm{t,1}}\geq p_{\textrm{r,1}}$; otherwise, $\omega=-\arctan((p_{\textrm{t,2}}-p_{\textrm{r,2}})/(p_{\textrm{t,1}}-p_{\textrm{r,1}}))-\pi$.

After the STN extracting the local scatter structures with rotation
and scaling invariance, the CNN takes the transformed scatter feature
map $\mathbf{F}$ as input to learn the scattering channel attenuation,
i.e., $g_{\mathrm{s}}(\tilde{\mathbf{p}},\mathbf{H})=f_{\mathrm{CNN}}(\mathbf{F})$.
The architecture of the proposed CNN is plotted in Fig.~\ref{fig:CNN}.
Two inception modules \cite{SzeChrWei:15C} are implemented in this
work to process and aggregate the scatter feature map at various scales,
so that the next stage can abstract higher-level features from the
different scales simultaneously for deeper scatter structure learning.
Any other CNN architecture is available in our work. Therefore, the
scattering branch combines the STN with CNN to learn the geometry
structure of the local environment by exploiting the rotation and
scaling invariance, which can improve the generalization ability of
the proposed network.

\subsection{Neural Network Training}

Given a set of \ac{rss} measurements $\{y^{(i)}\}$ taken at position
pair $\{\tilde{\mathbf{p}}^{(i)}\}$, the designed network is trained
with the mean-squared error (MSE) loss function as
\begin{equation}
\underset{\bm{\Theta},\mathbf{\mathbf{H}}}{\mbox{minimize}}\qquad\frac{1}{N}\sum_{i=1}^{N}(y^{(i)}-g(\tilde{\mathbf{p}}^{(i)},\bm{\Theta},\mathbf{H}))^{2}\label{eq:estimation-problem}
\end{equation}
where $\bm{\Theta}$ is a collection of all parameters for each component.
We implement the designed network using the deep learning library
Pytorch and the Adam optimizer is used.

To obtain an initial virtual obstacle map $\mathbf{H}$, a simplified
clustering algorithm is applied to classify the training dataset into
2 categories (i.e. LOS and NLOS). A new dataset with category labels,
i.e., $(\tilde{\mathbf{p}}^{(i)},0)$ for the LOS category and $(\tilde{\mathbf{p}}^{(i)},1)$
for the NLOS category, is constructed to train the virtual obstacle
map $\mathbf{H}$ by equation (\ref{eq:obstruction_level}), as a
binary classification task. The details of the initialization algorithm
are summarized in Algorithm \ref{Alg1:Init-obstacle-map}.
\begin{algorithm}
\begin{algorithmic}[1]

\STATE Input training dataset $(\tilde{\mathbf{p}}^{(i)},y^{(i)})$,
$i=0,1,...,N$

\STATE Initialize categories $\mathcal{C}_{0}$, $\mathcal{C}_{1}$
and parameters $\beta_{1},\gamma_{1},\beta_{2},\gamma_{2}$

\FOR{$n=1$ to $K$}

\STATE Empty $\mathcal{C}_{0}$ and $\mathcal{C}_{1}$

\FOR{$i=0$ to $N$}

\STATE $\hat{g}_{1}^{(i)}=\beta_{1}+\gamma_{1}\log_{10}\|\mathbf{p}_{\mathrm{t}}^{(i)}-\mathbf{p}_{\mathrm{r}}^{(i)}\|_{2}$

\STATE $\hat{g}_{2}^{(i)}=\beta_{2}+\gamma_{2}\log_{10}\|\mathbf{p}_{\mathrm{t}}^{(i)}-\mathbf{p}_{\mathrm{r}}^{(i)}\|_{2}$

\IF { $|y^{(i)}-\hat{g}_{1}^{(i)}|\leq|y^{(i)}-\hat{g}_{2}^{(i)}|$
}

\STATE $\tilde{\mathbf{p}}^{(i)}\rightarrow$ $\mathcal{C}_{0}$

\ELSE \STATE $\tilde{\mathbf{p}}^{(i)}\rightarrow$ $\mathcal{C}_{1}$

\ENDIF \ENDFOR

\STATE $\underset{\beta_{1},\gamma_{1}}{\mbox{Minimize}}||y^{(i)}-\hat{g}_{1}||$,
$\forall\tilde{\mathbf{p}}^{(i)}\in$ $\mathcal{C}_{0}$

$\underset{\beta_{2},\gamma_{2}}{\mbox{Minimize}}||y^{(i)}-\hat{g}_{2}||$,
$\forall\tilde{\mathbf{p}}^{(i)}\in$ $\mathcal{C}_{1}$

\STATE Update $\beta_{1},\gamma_{1},\beta_{2},\gamma_{2}$

\ENDFOR

\IF {mean$(\hat{g}_{1}^{(i)})\leq$ mean$(\hat{g}_{2}^{(i)})$ }

\STATE $\mathcal{C}_{0}$ is the LOS category (label with 0)

$\mathcal{C}_{1}$ is the NLOS category (label with 1)

\ELSE \STATE $\mathcal{C}_{1}$ is the LOS category; $\mathcal{C}_{0}$
is the NLOS category

\ENDIF \STATE Construct a new dataset $\{\tilde{\mathbf{p}}^{(i)},c^{(i)}\}$
where $c^{(i)}\in\{0,1\}$

\STATE Train $\mathbf{H}$ by equation (\ref{eq:obstruction_level})
using category labels as

$\underset{\mathbf{\mathbf{H}}}{\mbox{minimize}}-\frac{1}{N}\sum_{i=1}^{N}c^{(i)}\textrm{log}(I)+(1-c^{(i)})\textrm{log}(1-I)$

\end{algorithmic}

\caption{Learning an initial virtual obstacle map $\mathbf{H}$.}

\label{Alg1:Init-obstacle-map}
\end{algorithm}

\section{Simulations}

In this section, numerical results are presented to demonstrate the
effectiveness of the proposed model for joint radio map and virtual
obstacle map construction. To train and evaluate the proposed model,
we collect the data of channel measurements from a reference 3D RT
model.

In the simulation, we consider an air-to-ground communication scenario
in an urban environment of Shanghai and Beijing, China. The region
of interest is a square area of side 650 meters and the aerial TXs
are randomly distributed with altitudes ranging from 50 to 200 meters.
The heights of the ground RXs are fixed at 1.5 meters. The 3D geometry
of the surroundings is obtained from the recent city map database.
The channel measurements are generated using Remcom Wireless Insite,
with up to 6 reflections and 1 diffraction. The waveform is chosen
as sinusoidal signal at 5.9 GHz with 10MHz bandwidth and there is
3 dB noise to the RXs.

Based on the above setting, we construct three datasets from different
environments. Dataset I and Dataset II are generated from the city
areas in Shanghai and Beijing, China, respectively. There are $600$k
measurements at a TX altitude of 50 meters and nearly $190$k measurements
in the other altitudes for both datasets. Dataset III is referred
to as the \emph{diffraction dataset} that consists of only the NLOS
propagation scenarios where the position pair $\tilde{\mathbf{p}}$
is blocked by at least one obstacle. In this dataset, the channel
measurements are generated using Remcom Wireless Insite with the setting
of up to 3 diffraction and no reflection which contains $300$k measurements.

\begin{figure*}[!t]
\subfigure[]{ \includegraphics[scale=0.33]{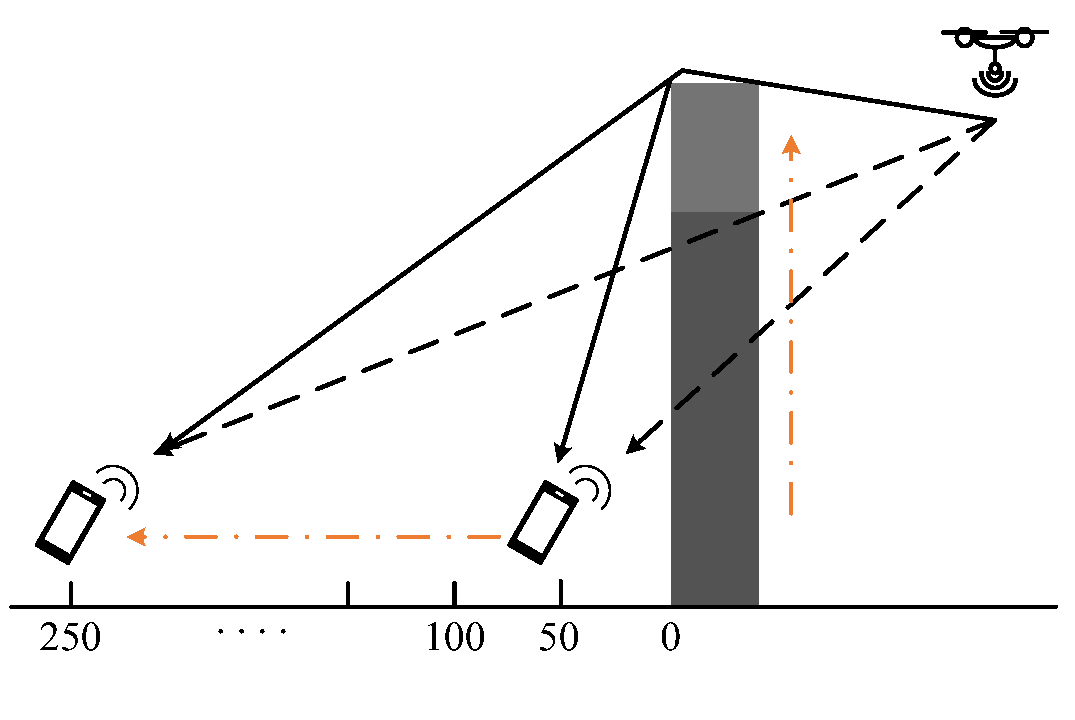}
}\subfigure[]{\label{fig:NLOS_example-11} \includegraphics[scale=0.4]{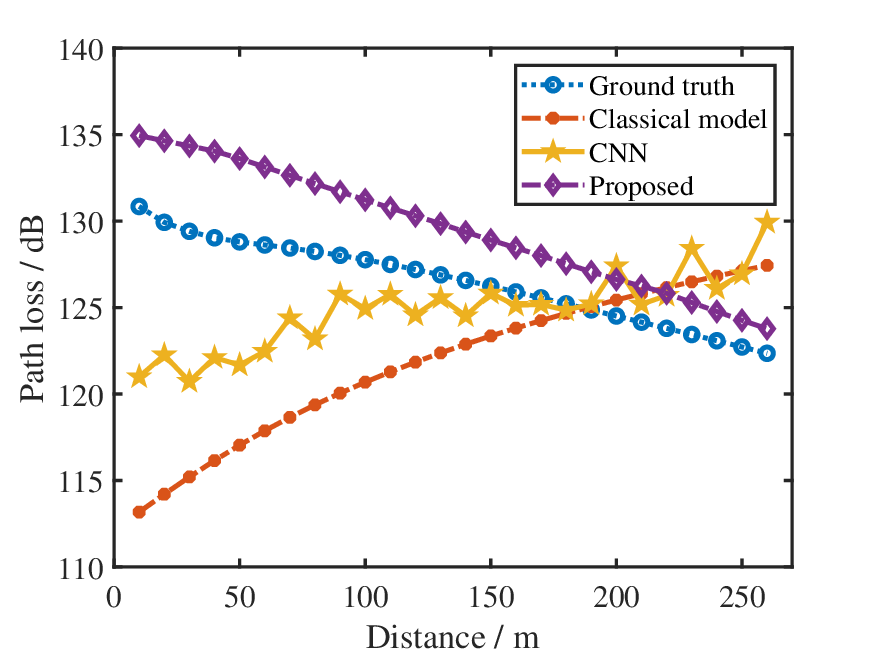}}\subfigure[]{\label{fig:NLOS_example-12}
\includegraphics[scale=0.4]{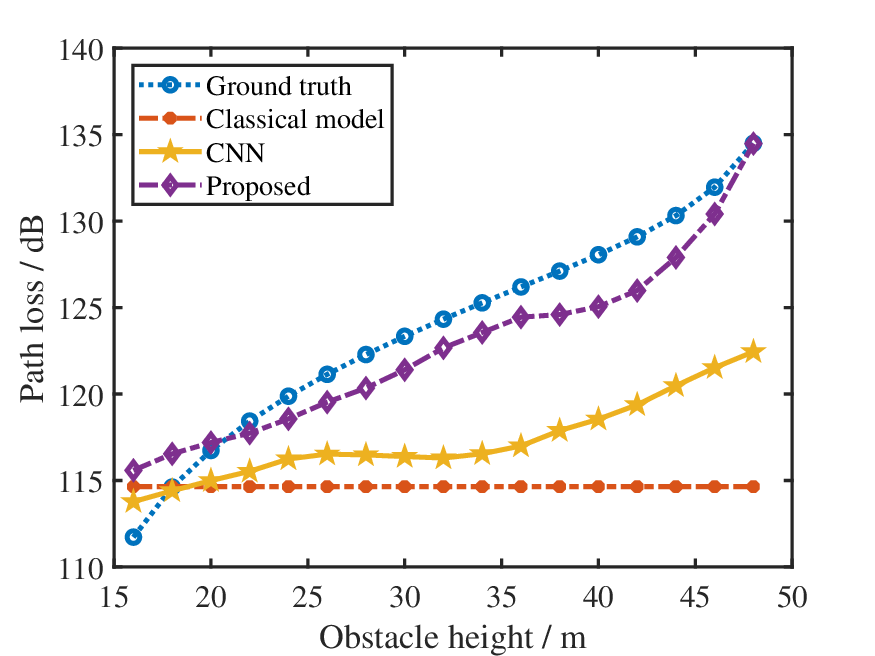}}\caption{a) A special example of the diffraction propagation scenario with
a single obstacle; b) With the RX moving away from the obstacle, the
path loss is decreasing as the increase of the distance due to the
diffraction effect; c) With the increase of the obstacle height, the
diffraction effect diminishes resulting in larger channel path loss.}
\label{fig:NLOS_example}
\end{figure*}

The proposed model is compared with the following baselines, MLP \cite{KenSaiChe:J19},
RadioUNet \cite{LevYapKut:J21}, PMNet \cite{JuOmeDhe:22C} and $k$-nearest
neighbor (KNN), which are summarized as follows:

1) MLP \cite{KenSaiChe:J19}: This method is a data-fitting multilayer
perceptron (MLP). The network partitions the samples into three clusters
and fit the data in each cluster using NN with seven dense layers.
It receives the TX-RX position pair $\tilde{\mathbf{p}}$ as input
and returns the estimation of the path loss.

2) RadioUNet \cite{LevYapKut:J21}: The network consists of two UNets.
Each consists of 9 encoder layers, and each layer consists of convolution,
ReLU, and Maxpool layer. Similarly, there are 9 decoder layers, each
consisting of transposed convolution followed by ReLU. The first UNet
takes the city map and the TX position as inputs and outputs a coarse
radio map. The second UNet takes the city map and the TX position
plus the coarse radio map to generate the final radio map.

3) PMNet \cite{JuOmeDhe:22C}: The network consists of 5 ResNet-based
encoder layers and 7 ResNet-based decoder layers. Each layer applies
several parallel atrous convolutions, Maxpooling, and ReLU. It takes
the city map and the TX position map as input to generate the path
loss radio map as well.

4) KNN: In KNN, the algorithm selects 6 measurement samples that are
closest to $\tilde{\mathbf{p}}$ from the training set and then forms
the neighbor set as $\mathcal{N}(\tilde{\mathbf{p}})$. The channel
path loss at $\tilde{\mathbf{p}}$ is determined by $g(\tilde{\mathbf{p}})=\mu^{-1}\sum_{i\in\mathcal{N}(\tilde{\mathbf{p}})}\omega(\tilde{\mathbf{p}},\tilde{\mathbf{p}}^{(i)})y^{(i)}$,
where $\omega(\tilde{\mathbf{p}},\tilde{\mathbf{p}}^{(i)})=\exp[-||\tilde{\mathbf{p}}-\tilde{\mathbf{p}}^{(i)}||_{2}^{2}(2s^{2})]$
with a properly chosen parameter $s=50$ meters and $\mu=\sum_{i\in\mathcal{N}(\tilde{\mathbf{p}})}\omega(\tilde{\mathbf{p}},\tilde{\mathbf{p}}^{(i)})$
is a normalizing factor.

\subsection{Performance of the Proposed Diffraction Component}

In this experiment, we separate the diffraction component $g_{\mathrm{d}}$
from the proposed model and train it individually using the diffraction
Dataset III. The diffraction dataset is split into three parts,
where the first part is used for training, the second part is used
for validation test and the last part is used for generalization test
in the cases of single obstacle and multiple obstacles. The environment
map is provided to train the proposed component $g_{\mathrm{d}}$
and a baseline CNN in this experiment. Here we consider a standard
CNN architecture consisting of convolution, ReLU, and Maxpool layers
with a comparable parameter quantity as the proposed model. Another
classical large-scale distance fitting model is also considered as
a baseline, i.e., $PL(d)=A+B\log_{10}(d)$, where $A$ and $B$ are
the path loss exponents. The mean absolute error (MAE) between the
estimated and the actual channel realization is considered as the
performance metric.

As an illustrative example, a special case of diffraction propagation
for the NLOS scenario with a single obstacle is presented at Fig.~\ref{fig:NLOS_example}.
In the first experiment, the positions of the obstacle and TX are
fixed. The RX moves away from the virtual obstacle. The prediction
curves of the path loss generated by the proposed model and the baselines
are plotted in Fig.~\ref{fig:NLOS_example-11}. The curve of the
proposed model approximates the ground truth while the CNN and the
distance fitting fails to capture the path loss variation in this
NLOS scenario. It is surprisedly observed that the path loss is decreasing
as the increase of the distance, which is contrary to the common large-scale
distance model. It is because for the RX further away from the obstacle,
the signal is less attenuated due to the diffraction effect.

In the second experiment, the positions of the TX and RX are fixed.
The prediction curves in terms of the increasing obstacle heights
are plotted in Fig.~\ref{fig:NLOS_example-12}. With the increase
of the obstacle height, the predicted path loss of both the proposed
model and the CNN increase as the true one, but the proposed model
is more accurate compared with the CNN. The classical model is insensitive
to the change of the obstacle height and fails to capture the diffraction
effect in the NLOS scenario. By contrast, the proposed model shows
a great superiority in the NLOS scenario by exploring the geometry
structures from the diffraction distances and angles.

\begin{table}[t]
\caption{Comparison results in the diffraction scenarios.}
\centering\label{Tab:Example_NlOS} \renewcommand\arraystretch{1.5}
\begin{tabular}{ p{2cm}<{\centering}| p{1.3cm}<{\centering}| c| c}
\hline

\multirow{2}{*}{Scheme}  &\multirow{2}{*}{Validation} &\multicolumn{2}{c}{Test} \\
\cline{3-4}
                         &                     & Single obstacle        & Multiple obstacles \\
\hline
Classical model        &8.98           &9.25           &9.22      \\
\hline

CNN                      &7.82           &5.55            &8.23      \\
\hline

Proposed             &\textbf{7.36}           &\textbf{5.01}            &\textbf{7.41}      \\
\hline

\end{tabular}
\end{table}

In the third experiment, we evaluate the MAE of the proposed model,
the CNN and the classical model for the path loss prediction in the
Dataset III, which are summarized in Table \ref{Tab:Example_NlOS}.
It can be observed that the proposed model can reduce the MAE by 1.5
dB and 0.5 dB compared with the classical model and the CNN, respectively.
In the test cases of the single obstacle or the multiple obstacles,
the prediction performance of the classical model deteriorates. By
contrast, the proposed model generates the best radio map performance,
especially in single obstacle scenarios.

\subsection{Performance Improvement by the Scattering Component}

\begin{table}
\caption{Comparison of radio map accuracy with different methods.}

\centering\label{Tab:Example_RadioMap_Accuracy} \renewcommand\arraystretch{1.5}
\begin{tabular}{ p{3.9cm}<{\centering}| p{1.8cm}<{\centering}| p{1.8cm}<{\centering}}
\hline

\multirow{2}{*}{Scheme}  &\multicolumn{2}{c}{NMAE} \\
\cline{2-3}
                                              & Dataset I        & Dataset II \\
\hline
MLP \cite{KenSaiChe:J19}                      & 0.0775           & 0.0765      \\
\hline

RadioUNet \cite{LevYapKut:J21}                & 0.0841            & 0.0831      \\
\hline

PMNet \cite{JuOmeDhe:22C}                     & 0.0782             & 0.0754     \\
\hline

KNN                   & 0.0779             & 0.0772     \\
\hline

\multirow{2}{*}{\makecell[c]{Proposed (full model) \\ ($g_\text{0} + g_\text{d} + g_\text{s}$)}} &\multirow{2}{*}{\textbf{0.0697}}    &\multirow{2}{*}{\textbf{0.0677}}   \\
 & &   \\
\hline

\multirow{2}{*}{\makecell[c]{Proposed (partial model) \\ ($g_\text{0} + g_\text{d}$)}}           &\multirow{2}{*}{0.0916}             &\multirow{2}{*}{0.0906}    \\
 & & \\
\hline

\multirow{2}{*}{\makecell[c]{Proposed (partial model) \\ ($g_\text{0} + g_\text{s}$)}}          &\multirow{2}{*}{0.0812}             &\multirow{2}{*}{0.0790}    \\
 & & \\
\hline

\end{tabular}
\end{table}

In this experiment, we evaluate the performance of the radio map construction
of the proposed model under different components, i.e., the LOS component
$g_{\mathrm{0}}$, the diffraction component $g_{\mathrm{d}}$ and
the scattering component $g_{\mathrm{s}}$, using Dataset I and Dataset
II. We fix the altitude of the TX at 50 meter in this experiment and
the accuracy of the radio map construction is assessed by the normalized
mean absolute error (NMAE). The radio map accuracy comparison of the
proposed full model $(g_{\mathrm{0}}+g_{\mathrm{d}}+g_{\mathrm{s}})$,
the proposed partial model $(g_{\mathrm{0}}+g_{\mathrm{d}})$ and
the proposed partial model $(g_{\mathrm{0}}+g_{\mathrm{s}})$ are
presented in Table~\ref{Tab:Example_RadioMap_Accuracy}. A performance
increase is observed with the proposed full model $(g_{\mathrm{0}}+g_{\mathrm{d}}+g_{\mathrm{s}})$,
as opposed to just utilizing the proposed partial model $(g_{\mathrm{0}}+g_{\mathrm{d}})$
or $(g_{\mathrm{0}}+g_{\mathrm{s}})$. The model without the scattering
component $g_{\mathrm{s}}$ has the worst performance, where an decrease
in the predictive NMAE is observed to be nearly $35\%$, which demonstrates
the significance of the proposed scattering component $g_{\mathrm{s}}$.
As a whole, each proposed component can provide a performance improvement
in the proposed model of radio map construction.

\subsection{Reconstruction of Virtual Obstacle Environment}

We further demonstrate the recovered geometry of the surroundings
that represent the radio propagation environment captured by the proposed
virtual obstacle map $\mathbf{H}$. We present the results of the
radio map construction in Dataset I.

\begin{figure}[!t]
\subfigure[]{\label{Obs_Map_true_map} \hspace{0.5cm}\includegraphics[width=3.5cm,height=3.5cm]{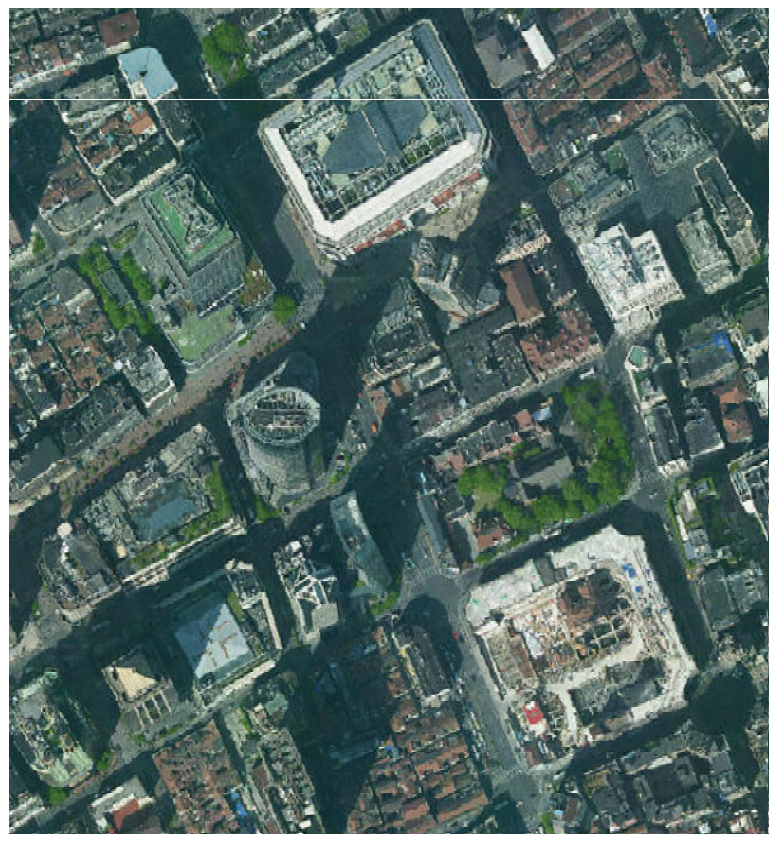}}
\subfigure[]{\label{Obs_Map_true_map_shape} \hspace{0.5cm}\includegraphics[width=3.65cm,height=3.5cm]{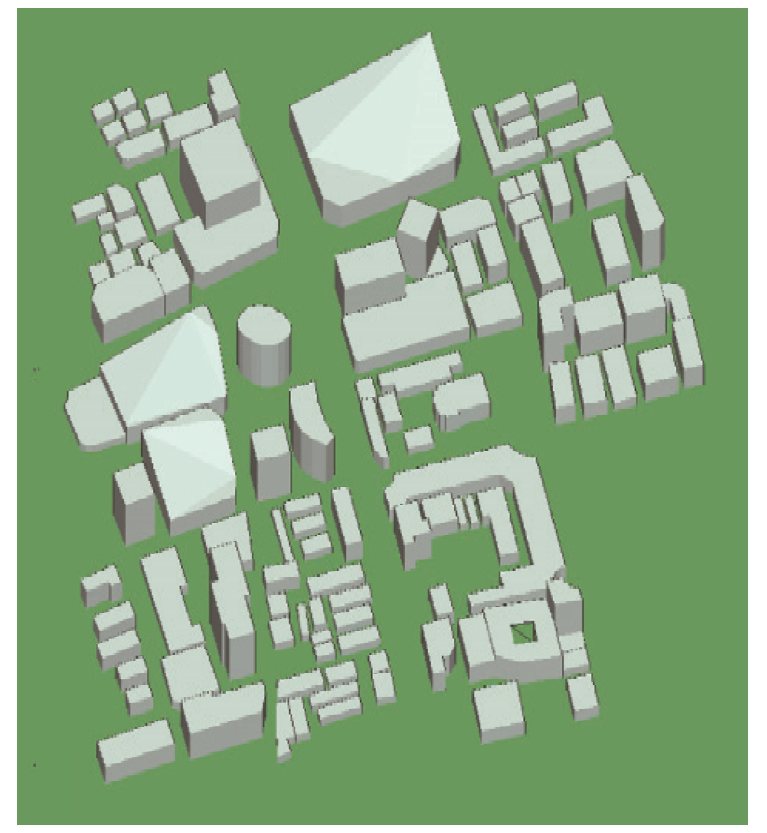}}

\subfigure[]{\label{Obs_Map_true_map_virtual1} \includegraphics[width=4cm,height=4cm]{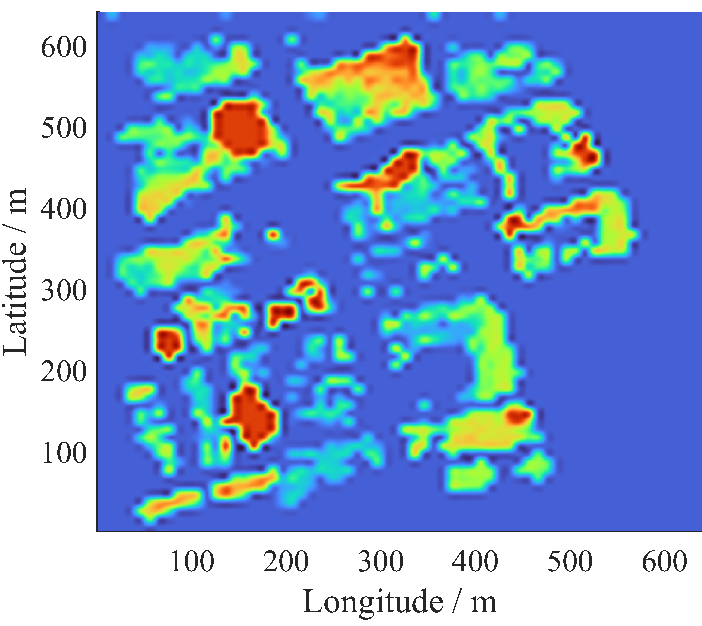}}
\subfigure[]{ \label{Obs_Map_true_map_virtual1-2}\includegraphics[width=4.6cm,height=4cm]{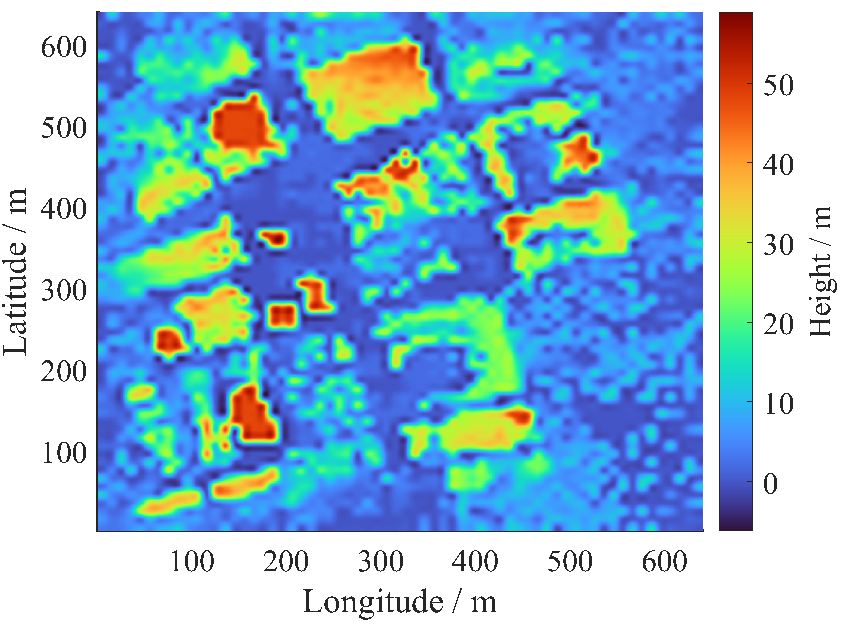}}

\caption{(a) The terrain map of an urban area of Nanjing Road in Shanghai;
(b) The 3D terrain of the city map; (c) The virtual obstacle map generated
by the proposed model without scattering component $g_{\mathrm{s}}$;
(b) The virtual obstacle map generated by the proposed model with
full components.}
\label{Obs_Map}
\end{figure}

The real city map of the Dataset I and its corresponding 3D terrain
model are presented in Fig.~\ref{Obs_Map_true_map} and Fig.~\ref{Obs_Map_true_map_shape}.
Fig.~\ref{Obs_Map_true_map_virtual1} shows the virtual obstacle
map $\mathbf{H}$ generated by the proposed model without the scattering
component $g_{\mathrm{s}}$ and Fig.~\ref{Obs_Map_true_map_virtual1-2}
is the virtual obstacle map $\mathbf{H}$ reconstructed by the proposed
full components. It can be seen that the geometry and distribution
of the virtual obstacles quite coincide with the true city map, which
can demonstrate the ability of the proposed model to uncover the hidden
environment information from the pure channel measurements. However,
it should be noted that the virtual obstacle map is not equivalent
to the true city map even though there is a morphological similarity
between them. Compared with Fig.~\ref{Obs_Map_true_map_virtual1}
and Fig.~\ref{Obs_Map_true_map_virtual1-2}, the virtual obstacle
map ignoring scattering seems more discriminative in terms of the
obstacle shapes and overall outline. With scattering, the virtual
obstacle map gets vaguer and unsmooth even in some obvious LOS regions.
It is because the virtual obstacle map servers as a geometry interpretation
of the radio propagation environment, where the scattering effect
incurs more complicated radio propagation characteristics and thus
leads to the complexity of radio environment.

The proposed virtual obstacle map represents the radio propagation
environment and provides blockage-aware ability of the channel condition.
Such ability can be used to predict the LOS/NLOS channel conditions
over the area of interest which has a great benefit in the radio map
construction and applications, which will be further discussed in
this paper.

\subsection{Comparison to Other State-of-the-arts}

We compare the performance of radio map construction of the proposed
model and the baselines in Dataset I and Dataset II. For a fair comparison,
we fix the altitude of the TX at 50 meters and NMAE is used to evaluate
the performance.

\begin{figure}[!t]
\centering \subfigure[]{ \includegraphics[width=3.9cm,height=4cm]{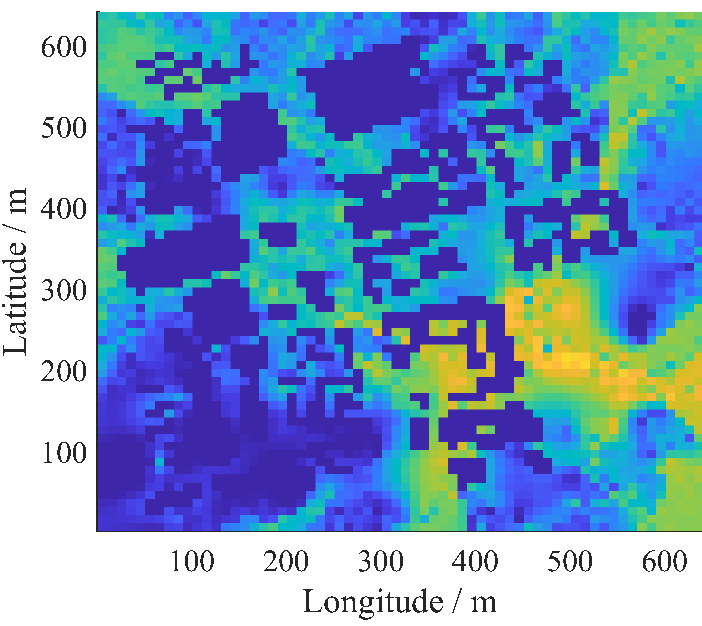}
}\subfigure[]{ \includegraphics[width=4.5cm,height=4cm]{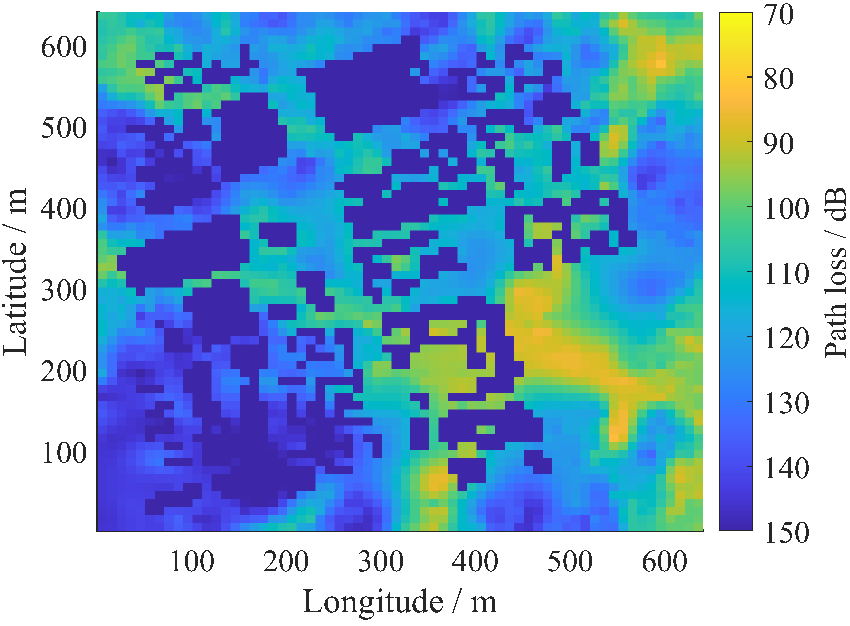}}\caption{A radio map of the path loss constructed from Dataset I at a fixed
TX position: a) The ground true radio map; b) The radio map generated
by the proposed model.}
\label{fig:RadioMap-Example}
\end{figure}

A 2D slice of the path loss radio map constructed by the proposed
model as well as the corresponding ground true radio map are presented
in Fig.~\ref{fig:RadioMap-Example}. It can be observed that the
radio geometry characteristics in a dense urban area can be very complicated
due to the distribution of a large number of obstacles. The radio
geometry of the proposed radio map is very closed to the ground true
one, which demonstrates that the proposed model is able to capture
the radio geometry characteristics in a complex environment for radio
map construction.

In Table~\ref{Tab:Example_RadioMap_Accuracy}, we also present the
NMAE of different design methods in terms of the radio map construction.
It is observed that the proposed model outperforms all the baselines
in both Dataset I and Dataset II, which offers a performance of $10\%$-$18\%$
accuracy improvements. In addition to accuracy improvements, the proposed
model does not require any city map compared with\emph{ }the RadioUNet
and PMNet, or a huge amounts of data to be stored compared with KNN,
which shows a great superiority in training and implementation.

\subsection{Transferability}

\begin{table*}[t]
\caption{Comparison results of radio map construction with transfer learning.}
\centering\label{Tab:Transferability-1}\renewcommand\arraystretch{1.5}
\begin{tabular}{ p{3cm}<{\centering}| p{2cm}<{\centering}| p{2cm}<{\centering}| p{2cm}<{\centering}| p{2cm}<{\centering}| p{2cm}<{\centering}| p{2cm}<{\centering}}
\hline

\multirow{3}{*}{Scheme}  &\multicolumn{3}{c|}{Transferred cases}  &\multicolumn{3}{c}{Non-transferred cases}\\
\cline{2-7}
                                              & $10\%$      & $30\%$    & $50\%$    & $10\%$      & $30\%$    & $50\%$ \\
\cline{2-7}
                                              &\multicolumn{6}{c}{Epoch / NMAE}   \\
\hline

RadioUNet \cite{LevYapKut:J21}                & 90 / 0.120       & 90 / 0.092    & 65 / 0.086    & 100 / 0.202
& 90 / 0.201    & 85 / 0.198 \\
\hline

PMNet \cite{JuOmeDhe:22C}                    & 100 / 0.107       & \textbf{97 / 0.082}    & \textbf{92 / 0.075}    & 100 / 0.132
& 100 / 0.107    & 91 / 0.106 \\
\hline

Proposed                                 & \textbf{50 / 0.106}       & 38 / 0.089    & \textbf{30 / 0.075}    & \textbf{100 / 0.125}      & \textbf{67 / 0.102}    & \textbf{50 / 0.081} \\
\hline

\end{tabular}
\end{table*}

\subsubsection{Generalize to Different Altitudes over 3D Space}

In this experiment, we demonstrate the proposed model is capable of
learning a high-dimensional radio map with full spatial degrees of
freedom where the TX can freely move over the 3D space as example.
In this experiment, the altitudes of the aerial TXs are fixed in 50
meters in the training phase while the altitudes will range from 50
to 200 meters in the test phase. The proposed model trained in 50
meters is generalized to predict the radio map in different altitudes
over the 3D space. We add two input channels to MLP model with TX
and RX heights, making it feasible for high-dimensional radio map
construction. The training data is from the Dataset I and the MAE
is used as the performance metric as well.

\begin{figure}[!t]
\centering \includegraphics[scale=0.5]{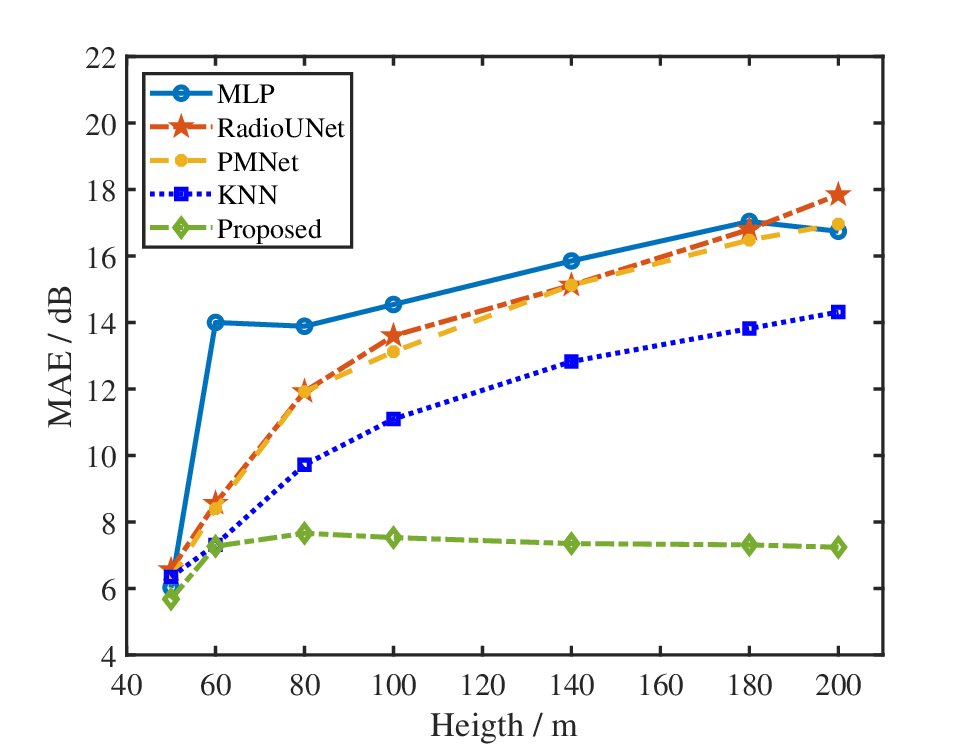}
\caption{Comparison results of the generalization to different TX altitudes
over the whole 3D space when the TX altitude in training phase is
fixed.}
\label{fig:RadioMap-with-heights}
\end{figure}

The comparison results of the radio map construction in different
TX altitudes in Dataset I are plotted in Fig.~\ref{fig:RadioMap-with-heights}.
It is obvious that the proposed model is superior to the baselines
in all TX altitudes over the 3D space. With the increase of the TX
altitudes, the MAE of the MLP, RadioUNet, PMNet and KNN increases
greatly. They cannot be generalized to predict the radio maps beyond
50 meters. By contrast, there are merely 1 dB deviation among all
altitudes in the proposed model without any additional measurement.
Therefore, the proposed model can be efficiently generalized to the
radio map construction with full spatial degrees of freedom without
any additional data to fine-tune the parameters, which have superiority
in the practical implementation.

\subsubsection{Transfer to a New Environment}

We further demonstrate the transfer ability of the proposed model
to a new environment. The notion of \emph{transfer learning} \cite{Goodfellow2016}
is applied as follows: first learn an initial parameters in an environment
and use the learned parameters as initialization for model learning
in a new environment. The impact of such initialization is that the
result of the model in the new environment will be generally closer
to a better local optimum than if a random initialization were adopted.

The proposed model is trained in Dataset I and transferred to the
Dataset II. In the case of transfer learning, the model is initialized
with the pre-trained parameters in Dataset I and different data volumes
with $10\%$, $30\%$ and $50\%$ of Dataset II are utilized to fine-tune
the model, respectively. For comparison, we also set up a case of
non-transfer learning, where the model is initialized randomly and
trained with the same data volumes of Dataset II. The proposed model
is compared with the RadioUNet and the PMNet in both cases of transfer
learning and non-transfer learning.

The comparison results of the training epoch and NMAE with/without
transfer learning are summarized in Table \ref{Tab:Transferability-1}.
It is observed that the transfer learning does improve the radio map
performance compared with the non-transfer learning in all cases.
With $50\%$ training data, the radio map accuracy of all schemes
under transfer learning approximates the optimal radio map accuracy
shown in Table.~\ref{Tab:Example_RadioMap_Accuracy}. By contrast,
the proposed method outperforms the baselines in all transferred/non-transferred
cases except the case of the transfer learning with $30\%$ training
data. Compared with the non-transferred cases, the proposed transferred
model converges faster with almost half of reduction in training epoch,
which shows a good transfer ability of the proposed model to a new
environment.

\subsection{Application of UAV-assisted Wireless Communication}

In this example, the superiority of the proposed model with the virtual
environment is demonstrated by using an application of UAV-assisted
wireless communication.

Consider to deploy a low altitude UAV to establish LOS channels for
two users in deep shadow in a dense urban area. Denote $\overline{g}(\mathbf{p},\mathbf{q})$
as the large-scale channel gain that captures both the path loss and
the shadowing between a terminal at position $\mathbf{p}$ and a terminal
at position $\mathbf{q}$. The objective of the UAV relay placement
problem is formulated as
\begin{align}
\mathop{\mbox{maximize}}\limits _{\mathbf{p}\in\mathbb{R}^{3}}\hspace{0.4cm} & \mathrm{min}\{\overline{g}(\mathbf{p},\mathbf{p}_{1}),\overline{g}(\mathbf{p},\mathbf{p}_{2})\}\label{eq:relay-placement-problem}\\
\text{subject to }\hspace{0.36cm} & (\mathbf{p},\mathbf{p}_{i})\in\mathcal{\tilde{D}}_{0},\qquad\qquad\forall i\in\{1,2\}.\nonumber
\end{align}
In the radio-map-based UAV relay placement problem, there is no any
additional information of environment except the radio map constructed
from the channel measurements. Because the classical radio-map-based
methods supply the pure radio map over an area of interest, only the
exhaustive 2D search or 3D search are available for UAV relay positioning.
By contrast, the proposed model can utilize the virtual environment
to provide blockage-aware ability of the channel condition which can
be adapted to a more intelligent searching strategy. Based on the
geometry property of the virtual obstacle model, if a UAV position
is double-LOS for two users, then all UAV positions perpendicularly
above it are double-LOS. The work \cite{ZheChe:22} has demonstrated
that it is empirically the best option to search on the middle perpendicular
plane of the two users for a nearly-optimal UAV relay position. The
UAV search strategy with the proposed model is summarized in two iterative
steps:
\begin{itemize}
\item Search downward when UAV is at double-LOS position;
\item Search along the circle with a fixed radius $r(\mathbf{p}')$ when
UAV is at non-double-LOS position, where $r(\mathbf{p}')=||\mathbf{p}'-\mathbf{o}||_{2}$
is the radius from a point $\mathbf{p}'$ on the perpendicular plane
to the midpoint of two users $\mathbf{o}$.
\end{itemize}
\begin{table}[t]
\centering \caption{Comparison results of the UAV relay placement.}
\label{tab:performance_known_user} \renewcommand\arraystretch{1.5}
\begin{tabular}{ p{2.5cm}<{\centering}| p{2.5cm}<{\centering}| p{2.5cm}<{\centering}}
\hline

Schemes           &Channel Gain (dB)       &Search Distance (km)      \\
\hline

Proposed          &-98.79                &1.371      \\
\hline

2D Search             &-100.91               &422.5      \\
\hline

3D Search             &-98.42              &6760      \\
\hline

MLP$_{3\text{D}}$             &-99.83               &6760      \\
\hline

\end{tabular}
\end{table}
The proposed UAV relay placement is compared with the exhaustive 2D
search (with fixed height at 50 meters) and exhaustive 3D search using
the same radio map generated by the proposed model. Another radio
map is generated by MLP \cite{KenSaiChe:J19} offline. Such radio
map also requires 3D search for the best UAV relay position, denoted
as $\textrm{\ensuremath{\textrm{MLP}_{\textrm{3D}}}}$.

The average optimized channel gain of 100 user position pairs as well
as the search distance generated by the proposed method, 2D search,
3D search and $\textrm{\ensuremath{\textrm{MLP}_{\textrm{3D}}}}$
are summarized in Table \ref{tab:performance_known_user}. It is observed
that the proposed method outperforms 2D search and $\textrm{\textrm{\ensuremath{\textrm{MLP}_{\textrm{3D}}}}}$
in terms of channel performance and search complexity. The 3D search
and $\textrm{\textrm{\ensuremath{\textrm{MLP}_{\textrm{3D}}}}}$ utilize
exhaustive search for the optimal UAV position with high expense,
but the performance of $\textrm{\textrm{\ensuremath{\textrm{MLP}_{\textrm{3D}}}}}$
is still worse than that of the proposed method due to the deficient
accuracy of the radio map. It demonstrates that the proposed virtual
environment is able to accelerate the UAV relay positioning. The performance
gap in channel gain compared with 3D search is within 0.5 dB, but
the proposed method shows great advantage in reducing more than $99\%$
search distance, which further demonstrates the efficiency and superiority
of the proposed method.

\section{Conclusion}

This paper developed a geometry model-assisted NN to explicitly exploit
the geometry structure of the environment for joint 6D radio map and
virtual obstacle map learning from the pure RSS data. In contrast
to many existing machine learning approaches that lack of an environment
model, a virtual obstacle model is developed to characterize the geometry
structure of the virtual environment. To capture the diffraction,
a transformer\emph{ }network is designed to mimic the computation
structure of the Vogler expression and learn the diffraction mechanism
from the key features of the diffraction distances and angles. In
addition, instead of directly mapping an entire city map to a radio
map, the proposed model only focuses on the geometry structure of
the local area surrounding the TX-RX pair to describe the scattering.
A CNN combined with a STN is designed to exploit the rotation invariance
and scale invariance of the local geometry structures for data augmentation
and scattering learning. Design examples show that the proposed model
can reconstruct the 3D geometry of the virtual environment and provide
$10\%-$$18\%$ accuracy improvement of the radio map construction.
It also shows a notable generalizability to the whole 3D space and
transferability to a new environment with $20\%$ training data reduction
and $50\%$ training epoch reduction. The usefulness is further illustrated
by an application of the radio-map-based UAV relay placement, where
the proposed model can reduce 99\% searching distance for UAV relay
positioning with the guarantee of an equivalent channel performance.

\begin{appendices}

\section{The Vogler Expression with Recursive Implementation \cite{NguPhaMan:J21}}\label{app:Vogler_method}

Consider the geometry of $N$ knife-edge diffraction, where $\{\theta_{i}\}_{n=1}^{N}$
and $\{d_{i}\}_{n=1}^{N}$ are $N$ diffraction angles and separation
distances between knife and edge, respectively. Then, the diffraction
attenuation $F_{N}$ for $N$ knife-edges is determined by
\begin{equation}
\begin{split}\mathop{F_{N}=} & \frac{1}{2^{N}}C_{N}e^{\sigma_{N}}\left(\frac{2}{\sqrt{\pi}}\right)\int_{\beta_{1}}^{\infty}\cdot\cdot\cdot\int_{\beta_{N}}^{\infty}\\
\textrm{} & e^{2F}\prod_{i=1}^{N}e^{-u_{N}^{2}}du_{1}\cdot\cdot\cdot du_{N}
\end{split}
\label{eq:diffraction-parameter}
\end{equation}
where
\begin{align*}
C_{N} & =\begin{cases}
1 & \textrm{for}\:N=1\\
\sqrt{\frac{\left(\sum_{i=1}^{N+1}d_{i}\right)\prod_{i=1}^{N}d_{i}}{\prod_{i=1}^{N}\left(d_{i}+d_{i+1}\right)}} & \textrm{for}\:N\geq2
\end{cases}\\
\sigma_{N} & =\sum_{i=1}^{N-1}\beta_{i}^{2}\\
F & =\begin{cases}
0 & \textrm{for}\:N=1\\
\sum_{i=1}^{N-1}\psi_{i} & \textrm{for}\:N\geq2
\end{cases}\\
\psi_{i} & =\phi_{i}(u_{i}-\beta_{i})(u_{i+1}-\beta_{i+1})\\
\beta_{i} & =\theta_{i}\left[\frac{\textrm{j}\pi d_{i}d_{i+1}}{\lambda(d_{i}+d_{i+1})}\right]^{1/2}\\
\phi_{i} & =\left[\frac{d_{i}d_{i+1}}{(d_{i}+d_{i+1})(d_{i+1}+d_{i+2})}\right]^{1/2}
\end{align*}
with $\lambda$ denoting the wavelength and $\textrm{j}=\sqrt{-1}$
being an imaginary number.

Consider $N>2$, by exploiting the fact that
\begin{equation}
\begin{split}\frac{2}{\sqrt{\pi}}\int_{\beta}^{\infty}(u-\beta)^{m}e^{-u^{2}}du=m!I(m,\beta)\end{split}
\label{eq:diffraction-parameter-1-2}
\end{equation}
where $m!$ refers to the factorial of $m$ and $I(m,\beta)$ defines
the repeated integrals of the complementary error function. A general
solution for $F_{N}$ is given as
\begin{equation}
\begin{split}\mathop{F_{N}=} & \frac{1}{2^{N}}C_{N}e^{\sigma_{N}}\sum_{m=0}^{\infty}I_{m}\end{split}
\label{eq:diffraction-parameter-1}
\end{equation}
where \begin{equation}
\resizebox{.98\hsize}{!}{$
\begin{split}I_{m} & =2^{m}\sum_{m_{1}=0}^{m}\cdots\sum_{m_{N-2}}^{m_{N-3}}\prod_{i=1}^{N}\frac{(m_{i-1}-m_{i+1})!}{(m_{i}-m_{i+1})!}\phi_{i}^{m_{i-1}-m_{i}}I(n_{i},\beta_{i})\end{split} \label{eq:diffraction-parameter-1-1}
$}
\end{equation}with, by using notion $m_{0}=m$
\begin{equation}
\begin{split}n_{i} & =\begin{cases}
m_{0}-m_{1}, & i=1\\
m_{i-2}-m_{i}, & 2\leq i\leq N-1\\
m_{N-2}-m_{N-1}, & i=N
\end{cases}.\end{split}
\label{eq:diffraction-parameter-1-1-1}
\end{equation}
Define
\begin{equation}
\begin{aligned}C(N-1 & ,m_{N-2},m_{N-3})\\
= & (m_{N-3})!\phi_{N-1}^{m_{N-2}}I(m_{N-3},\beta_{N-1})I(m_{N-2},\beta_{N}).
\end{aligned}
\label{eq:diffraction-parameter-1-1-1-1-1}
\end{equation}
Using the notion
\begin{equation}
\begin{aligned}i=m_{N-L}, & j=m_{N-L-1},k=m_{N-L-2}\\
2\leq L\leq N & -2,N\geq4
\end{aligned}
\label{eq:diffraction-parameter-1-1-1-1-2}
\end{equation}
and the recursive relationship
\begin{equation}
\begin{aligned}C(N & -L,j,k)\\
= & \sum_{i=0}^{j}\frac{(k-i)!}{(j-i)!}\phi_{N-L}^{j-i}I(k-i,\beta_{N-L})C(N-L+1,i,j)
\end{aligned}
\label{eq:diffraction-parameter-1-1-1-1-2-1}
\end{equation}
the recursive computation of $I_{m}$ is given as
\begin{equation}
\begin{split}I_{m} & =2^{m}\sum_{m_{1}=0}^{m_{0}}\phi_{1}^{m-m_{1}}I(m-m_{1},\beta_{1})\end{split}
C(2,m_{1},m).\label{eq:diffraction-parameter-1-1-1-1}
\end{equation}

Therefore, the computation of the Vogler expression can be represented
in a recursive manner, where the computation of $N$th diffraction
$F_{N}$ can depend on the computation of the $(N-1)$th diffraction
$F_{N-1}$. To emulate the computational structure inherent in the
Vogler expression, the utilization of dependencies and sequence natures
is paramount in the design of NN. The \emph{transformer} architecture
stands as an exemplar in this regard by harnessing the self-attention
mechanism to discern and model intricate relationships across the
sequence, which renders it particularly well-suited for tasks reliant
on understanding intricate dependencies and sequences, aligning with
the requisites of the computational structure of the Vogler expression.

\end{appendices}

\bibliographystyle{IEEEtran}
\bibliography{StringDefinitions,JCgroup,ChenBibCV,Bib/Reference,Bib/RadioMap,Bib/IEEEabrv}

\begin{thebibliography}{10}
\providecommand{\url}[1]{#1}
\csname url@samestyle\endcsname
\providecommand{\newblock}{\relax}
\providecommand{\bibinfo}[2]{#2}
\providecommand{\BIBentrySTDinterwordspacing}{\spaceskip=0pt\relax}
\providecommand{\BIBentryALTinterwordstretchfactor}{4}
\providecommand{\BIBentryALTinterwordspacing}{\spaceskip=\fontdimen2\font plus
\BIBentryALTinterwordstretchfactor\fontdimen3\font minus
  \fontdimen4\font\relax}
\providecommand{\BIBforeignlanguage}[2]{{%
\expandafter\ifx\csname l@#1\endcsname\relax
\typeout{** WARNING: IEEEtran.bst: No hyphenation pattern has been}%
\typeout{** loaded for the language `#1'. Using the pattern for}%
\typeout{** the default language instead.}%
\else
\language=\csname l@#1\endcsname
\fi
#2}}
\providecommand{\BIBdecl}{\relax}
\BIBdecl

\bibitem{MuLiuGuo:J21}
X.~Mu, Y.~Liu, L.~Guo, J.~Lin, and R.~Schober, ``Intelligent reflecting surface
  enhanced indoor robot path planning: A radio map-based approach,'' \emph{IEEE
  Trans. Wireless Commun.,}, vol.~20, no.~7, pp. 4732--4747, 2021.

\bibitem{WuYanXia:J17}
C.~Wu, Z.~Yang, and C.~Xiao, ``Automatic radio map adaptation for indoor
  localization using smartphones,'' \emph{IEEE Trans. on Mobile Computing},
  vol.~17, pp. 517--528, 2017.

\bibitem{ZenXu:J21}
Y.~Zeng and X.~Xu, ``Toward environment-aware {6G} communications via channel
  knowledge map,'' \emph{IEEE Wireless Commun.}, vol.~28, no.~3, pp. 84--91,
  2021.

\bibitem{HanXue:J20}
X.~Han, L.~Xue, Y.~Xu, and Z.~Liu, ``A two-phase transfer learning-based power
  spectrum maps reconstruction algorithm for underlay cognitive radio
  networks,'' \emph{IEEE Access}, vol.~8, pp. 81\,232--81\,245, 2020.

\bibitem{ZhaShuRui:J21}
S.~Zhang and R.~Zhang, ``Radio map-based {3D} path planning for
  cellular-connected {UAV},'' \emph{IEEE Trans. on Wireless Commun.,}, vol.~20,
  no.~3, pp. 1975--1989, 2021.

\bibitem{MoHuaXu:J20}
X.~Mo, Y.~Huang, and J.~Xu, ``Radio-map-based robust positioning optimization
  for {UAV}-enabled wireless power transfer,'' \emph{IEEE Wireless Commun.
  Lett.}, vol.~9, no.~2, pp. 179--183, 2020.

\bibitem{9354009}
Y.~Zeng, X.~Xu, S.~Jin, and R.~Zhang, ``Simultaneous navigation and radio
  mapping for cellular-connected {UAV} with deep reinforcement learning,''
  \emph{IEEE Trans. on Wireless Commun.}, vol.~20, no.~7, pp. 4205--4220, 2021.

\bibitem{9119191}
Q.~Hu, Y.~Cai, A.~Liu, G.~Yu, and G.~Y. Li, ``Low-complexity joint resource
  allocation and trajectory design for {UAV}-aided relay networks with the
  segmented ray-tracing channel model,'' \emph{IEEE Trans. on Wireless
  Commun.}, vol.~19, no.~9, pp. 6179--6195, 2020.

\bibitem{ZhuMaoSon:J22}
Q.~Zhu, K.~Mao, M.~Song, X.~Chen, B.~Hua, W.~Zhong, and X.~Ye, ``Map-based
  channel modeling and generation for {U2V} mm{W}ave communication,''
  \emph{IEEE Trans. Veh. Tech.}, vol.~71, no.~8, pp. 8004--8015, 2022.

\bibitem{LimChoSim:J20}
Y.-G. Lim, Y.~J. Cho, M.~S. Sim, Y.~Kim, C.-B. Chae, and R.~A. Valenzuela,
  ``Map-based millimeter-wave channel models: An overview, data for {B5G}
  evaluation and machine learning,'' \emph{IEEE Trans. Wireless Commun.},
  vol.~27, no.~4, pp. 54--62, 2020.

\bibitem{SugaSas:J21}
N.~Suga, R.~Sasaki, M.~Osawa, and T.~Furukawa, ``Ray tracing acceleration using
  total variation norm minimization for radio map simulation,'' \emph{IEEE
  Wireless Commun. Lett.}, vol.~10, no.~3, pp. 522--526, 2021.

\bibitem{SugaSas:J23}
N.~Suga, Y.~Maeda, and K.~Sato, ``Indoor radio map construction via ray tracing
  with {RGB-D} sensor-based {3D} reconstruction: Concept and experiments in
  {WLAN} systems,'' \emph{IEEE Access}, vol.~11, pp. 24\,863--24\,874, 2023.

\bibitem{ZhaWan:J22}
Y.~Zhang and S.~Wang, ``K-{N}earest neighbors {G}aussian process regression for
  urban radio map reconstruction,'' \emph{IEEE Commun. Lett.}, vol.~26, no.~12,
  pp. 3049--3053, 2022.

\bibitem{PhiTonSic:C12}
C.~Phillips, M.~Ton, D.~Sicker, and D.~Grunwald, ``Practical radio environment
  mapping with geostatistics,'' in \emph{Proc. IEEE Int. Symp. Dyn. Spectr.
  Access Netw.}, 2012, pp. 422--433.

\bibitem{SunChen:22J}
H.~Sun and J.~Chen, ``Propagation map reconstruction via interpolation assisted
  matrix completion,'' \emph{IEEE Trans. Signal Process.}, vol.~70, pp.
  6154--6169, 2022.

\bibitem{AusAndNev:J18}
A.~C.~M. Austin and M.~J. Neve, ``Efficient field reconstruction using
  compressive sensing,'' \emph{IEEE Trans. Antennas Propag.}, vol.~66, no.~3,
  pp. 1624--1627, 2018.

\bibitem{HuZha:J20}
Y.~Hu and R.~Zhang, ``A spatiotemporal approach for secure crowdsourced radio
  environment map construction,'' \emph{IEEE/ACM Trans. Netw.}, vol.~28, no.~4,
  pp. 1790--1803, 2020.

\bibitem{CheYatGes:C17}
J.~Chen, U.~Yatnalli, and D.~Gesbert, ``Learning radio maps for {UAV}-aided
  wireless networks: A segmented regression approach,'' in \emph{Proc. IEEE
  Int. Conf. Commun.}, Paris, France, May 2017.

\bibitem{KenSaiChe:J19}
K.~Saito, Y.~Jin, C.~Kang, J.~ichi Takada, and J.-S. Leu, ``Two-step path loss
  prediction by artificial neural network for wireless service area planning,''
  \emph{IEICE Commun. Exp.}, vol.~8, no.~12, pp. 611--616, 2019.

\bibitem{PopJefAta:J19}
S.~I. Popoola, A.~Jefia, A.~A. Atayero, O.~Kingsley, N.~Faruk, O.~F. Oseni, and
  R.~O. Abolade, ``Determination of neural network parameters for path loss
  prediction in very high frequency wireless channel,'' \emph{IEEE Access},
  vol.~7, pp. 150\,462--150\,483, 2019.

\bibitem{WuDanAi:20J}
L.~Wu, D.~He, B.~Ai, J.~Wang, H.~Qi, K.~Guan, and Z.~Zhong, ``Artificial neural
  network based path loss prediction for wireless communication network,''
  \emph{IEEE Access}, vol.~8, pp. 199\,523--199\,538, 2020.

\bibitem{LevYapKut:J21}
R.~Levie, Ã.~Yapar, G.~Kutyniok, and G.~Caire, ``Radio{UN}et: Fast radio map
  estimation with convolutional neural networks,'' \emph{IEEE Trans. Wireless
  Commun.}, vol.~20, no.~6, pp. 4001--4015, 2021.

\bibitem{JuOmeDhe:22C}
J.-H. Lee, O.~G. Serbetci, D.~P. Selvam, and A.~F. Molisch, ``{PMN}et: Robust
  pathloss map prediction via supervised learning,'' \emph{arXiv:2211.10527},
  2022.

\bibitem{AteHas:19J}
H.~F. Ates, S.~M. Hashir, T.~Baykas, and B.~K. Gunturk, ``Path loss exponent
  and shadowing factor prediction from satellite images using deep learning,''
  \emph{IEEE Access}, vol.~7, pp. 101\,366--101\,375, 2019.

\bibitem{ThrZibDar:20J}
J.~Thrane, D.~Zibar, and H.~L. Christiansen, ``Model-aided deep learning method
  for path loss prediction in mobile communication systems at 2.6 {GH}z,''
  \emph{IEEE Access}, vol.~8, pp. 7925--7936, 2020.

\bibitem{AhmOmaAte:20J}
O.~Ahmadien, H.~F. Ates, T.~Baykas, and B.~K. Gunturk, ``Predicting path loss
  distribution of an area from satellite images using deep learning,''
  \emph{IEEE Access}, vol.~8, pp. 64\,982--64\,991, 2020.

\bibitem{TegRom:J21}
Y.~Teganya and D.~Romero, ``Deep completion autoencoders for radio map
  estimation,'' \emph{IEEE Trans. Wireless Commun.}, vol.~21, no.~3, pp.
  1710--1724, 2022.

\bibitem{HuHua:J23}
T.~Hu, Y.~Huang, J.~Chen, Q.~Wu, and Z.~Gong, ``3{D} radio map reconstruction
  based on generative adversarial networks under constrained aircraft
  trajectories,'' \emph{IEEE Trans. Veh. Technol.}, vol.~72, no.~6, pp.
  8250--8255, 2023.

\bibitem{LiuChen:23J}
W.~Liu and J.~Chen, ``{UAV}-aided radio map construction exploiting environment
  semantics,'' \emph{IEEE Trans. on Wireless Commun.}, vol.~22, no.~9, pp.
  6341--6355, 2023.

\bibitem{ZenChe:C22}
P.~Zeng and J.~Chen, ``{UAV}-aided joint radio map and {3D} environment
  reconstruction using deep learning approaches,'' in \emph{Proc. IEEE Int.
  Conf. Commun.,}, 2022, pp. 5341--5346.

\bibitem{NguPhaMan:J21}
V.~D. Nguyen, H.~Phan, A.~Mansour, A.~Coatanhay, and T.~Marsault, ``On the
  proof of recursive vogler algorithm for multiple knife-edge diffraction,''
  \emph{IEEE Trans. Antennas Propag.}, vol.~69, no.~6, pp. 3617--3622, 2021.

\bibitem{vaswani:C2017}
A.~Vaswani, N.~Shazeer, N.~Parmar, J.~Uszkoreit, L.~Jones, A.~N. Gomez,
  L.~Kaiser, and I.~Polosukhin, ``Attention is all you need,'' in \emph{Proc.
  Adv. Neural Inf. Process. Syst. (NeurIPS)}, 2017, pp. 5998--6008.

\bibitem{JadSimZis:C15}
M.~Jaderberg, K.~Simonyan, A.~Zisserman, and k.~kavukcuoglu, ``Spatial
  transformer networks,'' in \emph{Proc. Adv. Neural Inf. Process. Syst.},
  vol.~28, 2015.

\bibitem{SzeChrWei:15C}
C.~Szegedy, W.~Liu, Y.~Jia, P.~Sermanet, S.~Reed, D.~Anguelov, D.~Erhan,
  V.~Vanhoucke, and A.~Rabinovich, ``Going deeper with convolutions,'' in
  \emph{Proc. IEEE Int. Conf. Comput. Vis. Pattern Recognit. (CVPR)}, 2015, pp.
  1--9.

\bibitem{Goodfellow2016}
I.~Goodfellow, Y.~Bengio, and A.~Courville, \emph{Deep Learning}.\hskip 1em
  plus 0.5em minus 0.4em\relax MIT Press, 2016.

\bibitem{ZheChe:22}
Y.~Zheng and J.~Chen, ``Geography-aware optimal {UAV} {3D} placement for {LOS}
  relaying: A geometry approach,'' \emph{arXiv:2209.15161}, 2022.

\end{thebibliography}

\end{document}